\pdfoutput=1
\documentclass[PAPER, atlasdraft=false, cernpreprint, UKenglish, texmf, orcidlogo]{atlasdoc}
\usepackage{atlaspackage}
\usepackage{atlasbiblatex}

\usepackage{atlasphysics}
\usepackage{xfrac} %
\usepackage{multirow}
\usepackage{makecell}

\graphicspath{{logos/}{figures/}}

\usepackage{ANA-EXOT-2022-05-PAPER-defs}

\AtlasTitle{Search for pair-produced vector-like $T$-quarks decaying into $Ht$ final states in the lepton-plus-jets channel in $pp$ collisions at $\sqrt{s}=13\,\mathrm{TeV}$ with the ATLAS detector}

\AtlasAbstract{%
A search for pair-produced vector-like $T$-quarks in 139 fb$^{-1}$ of $pp$ collisions recorded by the ATLAS detector at a centre-of-mass energy $\sqrt{s}=13\,\mathrm{TeV}$ is presented.
The search targets signal events where at least one $T$-quark decays into a Higgs boson and a top quark, $T\rightarrow Ht$, with subsequent decays of the Higgs boson into a bottom quark--antiquark pair, $H\rightarrow\bbbar$, and the top quark into a light charged lepton, a neutrino and a bottom quark, $t\rightarrow\ell^+\nu_\ell b$, where $\ell=e/\mu$.
The decay products of boosted heavy resonances are reconstructed using variable-radius, reclustered jets, and a neural network is employed to discriminate signal from Standard Model background processes.
A profile-likelihood fit to the effective mass distribution is performed, and no excess above the Standard Model prediction is observed.
Exclusion limits at $95\%$ confidence level are set on the $T$-quark pair-production cross-section, corresponding to lower mass limits of $1.40\,\mathrm{TeV}$ in the SU(2)~singlet representation, $1.56\,\mathrm{TeV}$ in the SU(2)~doublet representation, and $1.66\,\mathrm{TeV}$ assuming the branching fraction of $T$-quarks to $Ht$ to be $100\%$.
}

\AtlasRefCode{EXOT-2022-05}

\PreprintIdNumber{CERN-EP-2026-131}

\AtlasJournal{JHEP}
\AtlasCoverEgroupAnalysisTeam{atlas-exot-2022-05-analysis-team}

\hypersetup{pdftitle={ATLAS document},pdfauthor={The ATLAS Collaboration}}

\begin{document}

\maketitle

\tableofcontents

%
%

\section{Introduction}
\label{sec:intro}

One of the unsolved questions of particle physics is the large hierarchy between the electroweak mass scale and the Planck scale.
A possible way to alleviate the associated fine-tuning is to extend the Standard Model (SM) by introducing new, vector-like quarks (VLQs)~\cite{Dimopoulos:1981xc,Kaplan:1983fs,Kaplan:1983sm,Schmaltz:2005ky,Perelstein:2003wd,Gherghetta:2000qt,Martin:2009bg},
which can counterbalance the quadratically divergent radiative corrections to the Higgs boson mass.
VLQs are defined as colour-triplet spin-$\sfrac{1}{2}$ fermions whose left- and right-handed chiral components have the same transformation properties under the weak-isospin SU(2) gauge group~\cite{delAguila:1982fs,Aguilar-Saavedra:2013wba}.
The vector-like quarks in many of these models are expected to couple preferentially to third-generation SM quarks because of the large quark mass hierarchy in the SM~\cite{delAguila:1982fs,AguilarSaavedra:2009es}.
Several multiplet representations of vector-like quarks are possible, \eg \singlet{}s~$(T^{\sfrac{2}{3}})$ or \doublet{}s~$(X^{\sfrac{5}{3}},T^{\sfrac{2}{3}})$, where the superscript indicates the electric charge, which is omitted in the following.
The up-type \TQuark{}s are the focus of the search presented in this paper.
They can decay into a $W$ boson and a bottom quark~($Wb$), a Higgs boson and a top quark ($Ht$), or a $Z$ boson and a top quark ($Zt$).
The decay branching fractions are determined by the gauge representations of the vector-like quarks.
In the asymptotic limit of very large VLQ mass, these branching fractions converge to $50\% : 25\% : 25\%$ for $T\to Wb:Ht:Zt$ in the \singlet\ case, and $0:50\%:50\%$ in the \doublet\ case~\cite{Buchkremer:2013bha,AguilarSaavedra:2009es}, which are the values used in this analysis.

\begin{figure}
\centering
\subfloat[]{%
\includegraphics{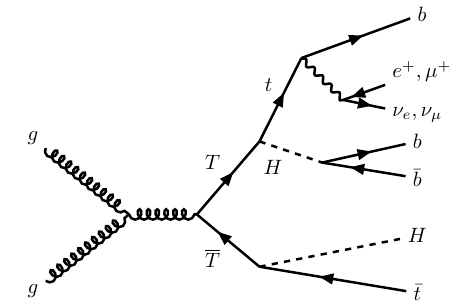}
}
\subfloat[]{%
\includegraphics{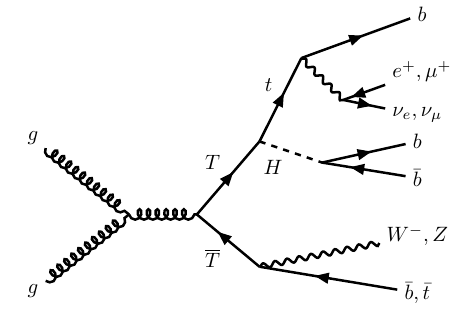}
}
\caption{%
Leading-order Feynman diagrams of the processes targeted by the search: a pair-production of vector-like \TQuark{}s, with at least one decaying through the $T\to Ht$ channel with subsequent decays $H\to\bbbar$ and $t\to\ell^+\nu_\ell b$, $\ell=e/\mu$.
The other \TQuark\ can either (a) decay as $T\to Ht$~(leading to final state~\HtHt) or (b) decay as $T\to Wb$~(\HtWb) or $T\to Zt$~(\HtZt), with subsequent hadronic decays of the bosons and top quark(s).
}
\label{fig:feynman_diagram}
\end{figure}

This paper presents a search for pair-produced \TQuark{}s in proton--proton collisions recorded at a centre-of-mass energy of $\sqrt{s}=\SI{13}{TeV}$ with the ATLAS detector~\cite{PERF-2007-01} at the Large Hadron Collider (LHC), using the full Run~2 data corresponding to an integrated luminosity of \usedIntLumiUnit.
The search targets events where at least one \TQuark\ decays into a Higgs boson and a top quark, $T\to Ht$, with a subsequent decay of the Higgs boson into a bottom quark--antiquark pair, $H\to\bbbar$, and a top quark into a light lepton, a neutrino and a bottom quark, $t\to\ell^+\nu_\ell b$, as depicted in Figure~\ref{fig:feynman_diagram}.
The second \TQuark\ is allowed to decay via any of the three decay modes ($T\to Ht,Zt,Wb$), followed by fully hadronic decay of these particles. %
Only light leptons, \ie electrons and muons, are targeted in this analysis.
The signal topology selected by this search is therefore characterised by the presence of exactly one electron or muon, moderate missing transverse momentum from neutrinos, and multiple boosted hadronic objects from the quarks or hadronically decaying bosons.
Explicit reconstruction of the signal decay products is combined with a neural network classifier to discriminate signal from background.
The search is performed in the \singlet\ and doublet gauge representations as well as a scenario where the branching fraction of \TQuark{}s to $Ht$ is set to $100\%$ (\BTtoHtOne).
The results are interpreted as upper limits on the vector-like \TQuark\ pair-production~(\TT) cross-section as a function of \TQuark\ mass.

The pair-production of \TQuark{}s was previously searched for by the ATLAS Collaboration using \SI{36}{\per\fb} of Run 2 data, both in the $T\to Ht$ channel~\cite{EXOT-2016-13} and in a combination of several decay channels~\cite{EXOT-2017-17}.
More recently, several searches for the pair-production of \TQuark{}s using the full Run 2 data were performed by the ATLAS Collaboration~\cite{EXOT-2018-58,EXOT-2019-08,EXOT-2019-06,EXOT-2021-35} and CMS Collaboration~\cite{CMS-B2G-20-011} in different decay channels.
To date, the most stringent limits exclude masses below \SI{1.48}{TeV} in the \singlet\ scenario, below \SI{1.49}{TeV} in the \doublet\ scenario, and below \SI{1.50}{TeV} in the \BTtoHtOne\ scenario, all set by the CMS search in Ref.~\cite{CMS-B2G-20-011}.

The search presented here improves upon the previous ATLAS search in the $T\to Ht$ channel~\cite{EXOT-2016-13} in several key aspects.
The modelling of the dominant \ttbar and vector-boson backgrounds is improved through a data-driven reweighting technique.
Explicit reconstruction of leptonic top-quark and $W$-boson candidates is performed, and the back-to-back topology characteristic of pair-produced heavy particles is exploited, providing additional handles for the separation of signal from background processes.
Finally, a neural network is employed for signal-to-background discrimination, capitalising on the almost four-fold increase in integrated luminosity and enabling a more effective exploitation of the data.


\section{ATLAS detector}
\label{sec:detector}

The ATLAS experiment~\cite{PERF-2007-01} at the LHC is a multipurpose particle detector with a forward--backward symmetric cylindrical geometry and a near \(4\pi\) coverage in solid angle.\footnote{\AtlasCoordFootnote}
It consists of an inner tracking detector surrounded by a thin superconducting solenoid providing a \qty{2}{\tesla} axial magnetic field, electromagnetic and hadronic calorimeters, and a muon spectrometer.
The inner tracking detector covers the pseudorapidity range \(|\eta| < 2.5\).
It consists of silicon pixel, silicon microstrip, and transition radiation tracking detectors.
Lead/liquid-argon (LAr) sampling calorimeters provide electromagnetic (EM) energy measurements with high granularity within the region \(|\eta|< 3.2\).
A steel/scintillator-tile hadronic calorimeter covers the central pseudorapidity range (\(|\eta| < 1.7\)).
The endcap and forward regions are instrumented with LAr calorimeters for EM and hadronic energy measurements up to \(|\eta| = 4.9\).
The muon spectrometer surrounds the calorimeters and is based on three large superconducting air-core toroidal magnets with eight coils each.
The field integral of the toroids ranges between \num{2.0} and \qty{6.0}{\tesla\metre} across most of the detector.
The muon spectrometer includes a system of precision tracking chambers up to \(|\eta| = 2.7\) and fast detectors for triggering up to \(|\eta| = 2.4\).
The luminosity is measured mainly by the LUCID--2~\cite{LUCID2} detector, which is located close to the beampipe.
A two-level trigger system is used to select events~\cite{TRIG-2016-01}.
The first-level trigger is implemented in hardware and uses a subset of the detector information to accept events at a rate close to \qty{100}{\kHz}.
This is followed by a software-based trigger that reduces the accepted rate of complete events to \qty{1.25}{\kHz} on average depending on the data-taking conditions.
A software suite~\cite{SOFT-2022-02} is used in data simulation, in the reconstruction and analysis of real and simulated data, in detector operations, and in the trigger and data acquisition systems of the experiment.


\section{Data and simulated event samples}
\label{sec:dataMC}

This search uses $pp$ collision data at $\sqrt{s} = \SI{13}{TeV}$ collected by the ATLAS experiment in 2015–2018 and corresponding to an integrated luminosity of \usedIntLumiUnit~\cite{DAPR-2021-01, ATLAS-CONF-2019-021, LUCID2}.
Events are only considered if they were recorded with a trigger selecting a single electron, a single muon, or based on the missing transverse momentum~\pTmiss\ (with magnitude~\met), under stable beam conditions when all detector subsystems were known to be in a good operating condition~\cite{DAPR-2018-01}.

Single-lepton triggers~\cite{TRIG-2018-01, TRIG-2018-05} requiring the lepton to be isolated in the tracking system and the calorimeter are combined in a logical OR with triggers with a higher transverse momentum (\pT) threshold without isolation requirements to achieve maximum efficiency.
Single-muon triggers with a \pT threshold of (20) \SI{26}{GeV} in (2015) 2016–2018 and isolation requirements, are combined with triggers with a \SI{50}{GeV} threshold with no isolation requirement.
A trigger with a \SI{60}{GeV} threshold but only relying on the barrel muon spectrometer was added for the 2017--2018 data-taking period.
The lowest \pT threshold for single-electron triggers was (24) \SI{26}{GeV} in (2015) 2016–2018.
The \met trigger~\cite{TRIG-2019-01} used an \met threshold of \SI{70}{GeV} in 2015 and a run-period-dependent \met threshold varying between \SI{90}{GeV} and \SI{110}{GeV} in other years.

The expected signal and background contributions are modelled by Monte Carlo (MC) simulated samples, with generators and configurations summarised in Table~\ref{tab:mc_summary}.
The simulated events were processed through a simulation~\cite{SOFT-2010-01} of the ATLAS detector geometry and response using \GEANT~\cite{Agostinelli:2002hh}, and reconstructed with the same software as the collision data.
All simulated samples, except those produced with \Sherpa~\cite{Hoeche:2009rj}, used \EvtGen[1.2.0]~\cite{Lange:2001uf} to model the heavy-flavour hadron decays.
The effect of multiple interactions in the same and neighbouring bunch crossings (pile-up) was modelled by overlaying minimum-bias interactions simulated with \Pythia[8.186]~\cite{Sjostrand:2007gs} using the A3 tune~\cite{ATL-PHYS-PUB-2011-014} onto the hard-scatter events according to the luminosity profile of the recorded data.
Corrections were applied to simulated events to ensure that lepton and jet selection efficiencies, energy scales, and energy resolutions match those measured in data.
The simulated samples were normalised to their respective cross-sections, calculated to the highest order
available in perturbation theory.
The top-quark and Higgs-boson masses were set to \SI{172.5}{GeV} and \SI{125}{GeV}, respectively, in all simulated samples.
Although the search directly targets only light leptons, $\tau$-lepton production and decay, in particular to electrons or muons, are included in all simulations.

\begin{table}[htbp]
\centering
\caption{
Summary of Monte Carlo samples used for event generation of signal and background processes.
The samples used to estimate the systematic uncertainties are indicated in parentheses.
$V$ denotes the production of an electroweak boson ($W$ or $Z$).
The column labelled ``Parton shower'' refers to the generator used for the parton shower~(PS), hadronisation and the underlying event.
The parton distribution function (PDF) used for the matrix element~(ME) is shown.
``Tune'' refers to the underlying set of tuned parameters for the PS generator.
\MGaMCshort\ refers to \MGNLO[2.2], 2.3 or 2.6~\cite{Alwall:2014hca}.
All samples include leading-logarithm photon emission, either modelled by the PS generator
or by \PHOTOS~\cite{Golonka:2005pn}.
}
\label{tab:mc_summary}
\renewcommand{\arraystretch}{1.2} %
\resizebox{\textwidth}{!}{
\begin{tabular}{llllll}
\toprule
\textbf{Process} & \textbf{Generator} & \textbf{ME order} & \textbf{Parton shower} & \textbf{PDF} & \textbf{Tune} \\
\midrule
\multicolumn{6}{c}{\textit{Signal}} \\
\midrule
\TT & \PROTOS[2.2]~\cite{protos} & \LO & \Pythia[8.1]~\cite{Sjostrand:2007gs} & \NNPDF[2.3lo]~\cite{Ball:2012cx} & A14~\cite{ATL-PHYS-PUB-2014-021} \\
\midrule
\multicolumn{6}{c}{\textit{Top-quark pairs}} \\
\midrule
\ttbar\ & \POWHEGBOX~\cite{Nason:2004rx,Frixione:2007vw,Alioli:2010xd} & \NLO & \Pythia[8.2]~\cite{Sjostrand:2014zea} & \NNPDF[3.0nlo]~\cite{Ball:2014uwa} & A14 \\
& (\MGaMCshort~\cite{Alwall:2014hca}) & (\NLO) & (\Pythia[8.2]) & (\NNPDF[3.0nlo]) & (A14) \\
& (\POWHEGBOX) & (\NLO) & (\Herwig[7.1.3]~\cite{Bahr:2008pv}) & (\NNPDF[3.0nlo]) & (\Herwig default) \\
($\ttbar\bbbar$) & (\POWHEGBOXRES~\cite{Jezo:2018yaf,ttbbPowheg}) & (\NLO) & (\Pythia[8.3]~\cite{Bierlich:2022pfr}) & (\NNPDF[3.1nnlo]~\cite{NNPDF:2017mvq}) & (A14) \\
\midrule
\multicolumn{6}{c}{\textit{Other top}} \\
\midrule
Single-top & \POWHEGBOX & \NLO & \Pythia[8.2] & \NNPDF[3.0nlo] & A14 \\
& (\POWHEGBOX) & (\NLO) & (\Pythia[8.2]) & (\NNPDF[3.0nlo]) & (A14) \\
& (\MGaMCshort) & (\NLO) & (\Pythia[8.2]) & (\NNPDF[3.0nlo]) & (A14) \\
& (\POWHEGBOX) & (\NLO) & (\Herwig[7.0.4]) & (\NNPDF[3.0nlo]) & (\Herwig default) \\
$\ttbar W$, \ttZy & \MGaMCshort & \NLO & \Pythia[8.2] & \NNPDF[3.0nlo] & A14 \\
$\ttbar H$ & \POWHEGBOX & \NLO & \Pythia[8.2] & \NNPDF[3.0nlo] & A14 \\
$tZ$ & \MGaMCshort & \LO & \Pythia[8.1] & \NNPDF[2.3lo] & A14 \\
$tH$ & \MGaMCshort & \NLO & \Pythia[8.2] & \NNPDF[3.0nlo] & A14 \\
$\ttbar\ttbar$ & \MGaMCshort & \LO & \Pythia[8.1] & \NNPDF[2.3lo] & A14 \\
\midrule
\multicolumn{6}{c}{\textit{Non-top}} \\
\midrule
\Vjets\ & \Sherpa[2.2.11]~\cite{Hoeche:2009rj} & \NLO/\LO & \Sherpa[2.2.11]~\cite{Schumann:2007mg} & \NNPDF[3.0nnlo] & \Sherpa\ default \\
$VV$ & \Sherpa[2.2.1/2.2.2] & \NLO/\LO & \Sherpa[2.2.1/2.2.2] & \NNPDF[3.0nnlo] & \Sherpa\ default \\
Multijet & \Pythia[8.2] & \LO & \Pythia[8.2] & \NNPDF[2.3lo] & A14 \\
\bottomrule
\end{tabular}
}
\end{table}


\subsection{Simulated signal samples}

Vector-like \TQuark\ pair-production signal samples were simulated at leading-order (\LO) with \PROTOS[2.2]~\cite{protos} using the \NNPDF[2.3lo]~\cite{Ball:2012cx} parton distribution function (PDF) set, interfaced to \Pythia[8.186] for parton showering and fragmentation with the A14~tune~\cite{ATL-PHYS-PUB-2014-021}.
The theoretical \TT\ cross-section was computed at approximate next-to-next-to-leading-order (\NNLO) in quantum chromodynamics (QCD) with next-to-next-to-leading logarithmic (NNLL) soft-gluon resummation using \TOPpp[2.0]~\cite{Cacciari:2011hy, Beneke:2011mq, Baernreuther:2012ws, Czakon:2012zr, Czakon:2013goa,Czakon:2011xx} and the \MSTW \NNLO~\cite{Martin:2009iq, Martin:2009bu} PDF set.
The vector-like quarks were forced to decay with equal branching fractions of $\sfrac{1}{3}$ to each of the three decay modes ($Wb$, $Zt$, $Ht$), and other sets of branching fractions consistent with the three benchmark scenarios (\singlet, \doublet, \BTtoHtOne) were obtained by reweighting.
Samples were simulated assuming \singlet\ couplings for \TQuark{} masses between \SI{0.8}{TeV} and \SI{2.0}{TeV}, with \SI{0.1}{TeV} intervals between \SI{1.0}{TeV} and \SI{1.8}{TeV}, corresponding to theoretical cross-sections between \SI[separate-uncertainty=true]{0.195\pm0.020}{pb} at \SI{0.8}{TeV} and \SI[separate-uncertainty=true]{0.132\pm0.024}{fb} at \SI{2.0}{TeV}.
Kinematic differences arising from the different chirality of \doublet\ and singlet couplings are found to be negligible, rendering the simulation of dedicated samples for the doublet scenario unnecessary.

\subsection{Simulated background samples}
\label{sec:bkgdSamples}
To suppress QCD and electroweak background processes that do not resemble the targeted signal topology, a series of preselection requirements are applied (see Section~\ref{sec:analysis} for details).
After preselection, the main source of background is the production of top quark--antiquark pairs ($t\bar{t}$), either alone or in association with jets.
Moreover, the production of a single top quark in association with a $W$~boson~($Wt$) contributes significantly in regimes of high transverse momentum.
It is grouped into the ``Other top'' category when not shown separately.
This category also includes subdominant background processes involving top quarks produced in association with bosons or additional top quarks, namely $\ttbar W, \ttZy, \ttbar H, tH, tZ$, and four-top events.
The remaining contributions originate mostly from \Wjets, \Zjets\ and QCD multijet production processes, which are grouped into the ``Non-top'' category when not shown separately.

Standard Model \ttbar events were produced in the five-flavour scheme at next-to-leading-order (\NLO) with \POWHEGBOX[v2]~\cite{Nason:2004rx,Frixione:2007vw,Alioli:2010xd,Frixione:2007nw} using the \NNPDF[3.0nlo]~\cite{Ball:2014uwa} PDF set.
Parton showering, hadronisation and
the underlying event were simulated using \Pythia[8.230]~\cite{Sjostrand:2014zea} with the \NNPDF[2.3lo] PDF set and the A14~tune.
This sample was normalised to the integrated cross-section calculated by \TOPpp[2.0] at \NNLO in QCD, including resummation of
NNLL soft-gluon terms.
The \hdamp~parameter in \POWHEGBOX, which controls the matching of the matrix element to the parton shower and effectively regulates the degree of high-\pT radiation, was set to $1.5\,m_t$~\cite{ATL-PHYS-PUB-2017-007}.
Two alternative \ttbar samples are used to estimate the effects of applying different matching prescriptions: one where the matrix element was calculated by \MGNLO[2.6.0]~\cite{Alwall:2014hca} instead of \POWHEGBOX, and one where showering and hadronisation was performed by \Herwig[7.1.3]~\cite{Bahr:2008pv} instead of \Pythia.

The \ttbar samples were simulated inclusively, with events categorised according to the flavour content of additional jets not originating from the decay of the \ttbar system (see Ref.~\cite{HIGG-2013-27} for details).
Events containing at least one \bHadron{} are classified as \ttbin.
Among the remaining events, those containing at least one \cHadron{} are classified as \ttcin.
All other events, including those with no additional jets, are labelled as \ttlight.
An alternative \ttbb\ sample was simulated at \NLO\ in the four-flavour scheme with \POWHEGBOXRES~\cite{Jezo:2018yaf,ttbbPowheg} interfaced to \OPENLOOPS~\cite{Buccioni:2019sur,Cascioli:2011va,Denner:2016kdg} and \Pythia[8.309]~\cite{Bierlich:2022pfr}, with the \NNPDF[3.1nnlo]~\cite{NNPDF:2017mvq} PDF set.
Based on the studies of Ref.~\cite{ATL-PHYS-PUB-2022-006}, the factorisation scale was set to $\frac{1}{2}\sum_{i=t,\bar{t},b,\bar{b},j} m_{\mathrm{T},i}$ and the renormalisation scale to $\frac{1}{2}\sqrt[4]{m_\mathrm{T}(t)\cdot m_\mathrm{T}(\bar{t})\cdot m_\mathrm{T}(b)\cdot m_\mathrm{T}(\bar{b})}$.
The \hdamp~parameter was set to $H_\mathrm{T}/2$ and the \POWHEG internal parameter $h_\mathrm{bzd}$, which regulates the damping function in conjunction with \hdamp, was set to 5.
This sample is used to assess the effects of alternative \ttbin\ modelling, in particular the use of the four- rather than the five-flavour scheme and consequently \bbbar pairs being included in the matrix element rather than being added by the parton shower.

Single-top-quark event samples for the $Wt$-, $s$-, and $t$-channel production mechanisms were simulated at \NLO\ with \POWHEGBOX[v2] using a fixed renormalisation and factorisation scale and the \NNPDF[3.0nlo] PDF set, interfaced to \Pythia[8.230] with the A14~tune and \NNPDF[2.3lo] PDF set.
Overlaps between the \ttbar and $Wt$ final states were removed using the diagram removal scheme~\cite{Frixione:2005vw, Frixione:2008yi}.
To assess the impact of this choice, an alternative sample was produced applying the diagram subtraction scheme but otherwise identical event settings.
Two alternative samples are used to estimate the effects of hard-process and parton-shower modelling, respectively: one where the matrix element was calculated by \MGNLO[2.6.2] instead of \POWHEGBOX, and one where showering and hadronisation was performed by \Herwig[7.0.4] instead of \Pythia.

The production of $\ttbar W$ and \ttZy\ events was modelled using \MGNLO[2.3.3] interfaced to \Pythia[8.210] in the A14~tune.
The production of $\ttbar H$ ($tH$) events was modelled using \POWHEGBOX[v2] (\MGNLO[2.3.3]) interfaced to \Pythia[8.230] in the A14~tune.
All of these samples were simulated at \NLO and using the \NNPDF[3.0nlo] PDF set.
The production of $tZ$ events was modelled at \LO using \MGNLO[2.2.1] with the \NNPDF[2.3lo] PDF set, interfaced to \Pythia[8.201] with the A14~tune.
Four-top-quark production ($\ttbar\ttbar$) was simulated at \LO using \MGNLO[2.2.2] with the \NNPDF[2.3lo] PDF set, interfaced to \Pythia[8.186] with the A14~tune, and normalised to the \NLO theoretical cross-section.

Samples of \Wjets\ and \Zjets\ events (\Vjets) were simulated using \Sherpa[2.2.11], following the ATLAS \Vjets\ modelling setup described in Ref.~\cite{PMGR-2021-01}.
The matrix-element calculation was performed with up to two partons at \NLO and up to five partons at \LO using \COMIX~\cite{Gleisberg:2008fv} and \OPENLOOPS.
The matrix-element calculation, performed using the \NNPDF[3.0nnlo] PDF set, was merged with the
\Sherpa parton shower~\cite{Schumann:2007mg} using the \MEPSatNLO prescription~\cite{Hoeche:2012yf}.

Samples of diboson final states ($VV$), representing collectively $WW$, $WZ$, and $ZZ$ samples, were simulated with \Sherpa[2.2.1] or 2.2.2 depending on the process, including off-shell effects and Higgs-boson contributions where appropriate.
Fully leptonic final states and semileptonic final states, where one boson decays leptonically and the other hadronically, were simulated at \NLO\ in QCD for up to one additional parton and at \LO\ for up to three additional parton emissions.
Samples for the loop-induced processes $gg\rightarrow VV$ were simulated at \LO\ for up to one additional parton emission for both, fully leptonic and semileptonic, final states.
The matrix-element calculations and PDF choice were handled using the same procedure as in \Vjets\ samples.

Multijet events were simulated at \LO\ using \Pythia[8.230] for dijet production and interfaced to a \pT-ordered parton shower~\cite{ATL-PHYS-PUB-2019-017}.


\section{Object reconstruction}
\label{sec:obj_reco}

The targeted final state, depicted in Figure~\ref{fig:feynman_diagram}, is characterised by exactly one electron or muon, multiple jets and $b$-tagged jets, and \met from the neutrinos.
Hadronically decaying vector bosons, Higgs bosons and top quarks are reconstructed by reclustering and tagging jets.
Leptonic $W$-boson candidates are constructed from the lepton and \pTmiss.
Leptonic top-quark candidates are then formed by combining the leptonic $W$-boson candidate with a $b$-tagged jet.
\TQuark\ candidates are constructed from reclustered jets and leptonic $W$-boson or top-quark candidates.
The kinematics of all objects are used to construct sensitive input variables for the multivariate signal classifier and to define analysis regions.
The detailed procedure for reconstructing all of these objects is described in the following.

Interaction vertices from proton--proton collisions are reconstructed from at least two tracks with $\pT>\SI{500}{MeV}$ that are consistent with originating from the beam collision region in the $x$--$y$ plane.
If more than one primary vertex candidate is found, the candidate whose associated tracks form the largest sum of squared \pT~\cite{ATL-PHYS-PUB-2015-026} is selected as the hard-scatter primary vertex.

Electron candidates~\cite{EGAM-2018-01} are reconstructed from energy clusters in the EM calorimeter associated with reconstructed tracks in the inner detector.
They are required to have $\pT>\SI{30}{GeV}$ and $|\eta_{\textrm{cluster}}| < 2.47$, omitting the transition region between the EM barrel and endcap calorimeters
($1.37 < |\eta_{\textrm{cluster}}| < 1.52$).
The ``Tight'' likelihood-based identification criteria using calorimeter, tracking, and combined variables are imposed to provide separation between electrons and jets.
To further reduce the background from heavy-flavour decays and misidentified light hadrons, the electron candidates are required to be isolated in the tracking system and the calorimeter according to the ``Loose'' criteria.

Muon candidates~\cite{MUON-2018-03} are reconstructed by matching tracks in the muon spectrometer to those found in the inner detector.
The resulting muon candidates are re-fitted using the complete track information from both detector systems.
They are required to have $\pT>\SI{30}{GeV}$, $|\eta|<2.5$, and satisfy the ``Medium'' (``High-$p_\mathrm{T}$'') identification criteria if their transverse momentum is smaller (larger) than \SI{800}{GeV}.
To further reduce the background from non-prompt muons, the muon candidates are required to be isolated in the tracking system and the calorimeter according to the ``TightTrackOnly'' criteria.
Electron (muon) candidates are matched to the primary vertex by requiring that the significance of their transverse impact parameter, $d_0$, satisfies $|d_0/\sigma(d_0)|<5\ (3)$, where $\sigma(d_0)$ is the measured uncertainty in $d_0$, and by requiring that their longitudinal impact parameter, $z_0$, satisfies $|z_0 \sin\theta|<\SI{0.5}{mm}$.

Jet candidates are reconstructed using the \antiKt\ algorithm~\cite{Cacciari:2008gp} as implemented in \FastJet~\cite{Fastjet} with a jet radius parameter $R=0.4$.
The inputs to this algorithm are particle-flow objects~\cite{PERF-2015-09}, which combine measurements from the ATLAS inner detector and calorimeters~\cite{PERF-2014-07}.
Candidate jets are required to have $\pT>\SI{25}{GeV}$ and $|\eta|<2.5$, and are called \smallR\ jets in the following.
Quality criteria are imposed to reject events that contain any jet arising from non-collision sources or detector noise~\cite{ATLAS-CONF-2015-029}.
For jets with $\pT<\SI{120}{GeV}$, the output score of the jet-vertex tagger (JVT)~\cite{PERF-2014-03} algorithm is required to be larger than 0.59 if $|\eta|<2.4$ and larger than 0.11 if $2.4\leq|\eta|<2.5$ to suppress contamination from jets that originate from pile-up interactions.

Overlaps between candidate objects are removed sequentially.
To suppress misattribution of muon calorimeter deposits, \smallR\ jets within $\Delta R = 0.4$ of a muon candidate are discarded if the muon has a significant fraction of the jet momentum or there is significant hadronic activity close to the muon.
Similarly, electrons that share a track with a muon candidate are removed.
Double-counting of clusters associated with electrons and \smallR\ jets is avoided by discarding \smallR\ jets within $\Delta R=0.2$ of any electron.
If the electron is within $\Delta R=0.4$ of the axis of any \smallR\ jet after this initial removal, the \smallR\ jet is retained and the electron is removed.
Jets arising from muon bremsstrahlung are characterised by having few matching inner-detector tracks.
They are removed by discarding those \smallR\ jet candidates within $\Delta R=0.2$ of a muon that have less than three tracks originating from the primary vertex.
Muons that satisfy $\Delta R(\mu,{\textrm{jet}}) < 0.04+\SI{10}{GeV}/\pT^\mu$ are rejected to remove those muons that are likely to have arisen in the decay chain of hadrons.

Jets originating from \bHadron{}s are identified using \antiKt\ variable-radius jets~\cite{Krohn:2009zg}, reconstructed from inner-detector tracks, with an effective radius parameter $\REff=\SI{30}{GeV}/\pT$ using the transverse momentum of the resulting jet, and minimum and maximum radius bounds of $0.02$ and $0.4$, respectively.
Flavour tagging is performed with the multivariate DL1r algorithm~\cite{FTAG-2019-07}, which exploits the jet kinematics, track-level impact parameter information, and the properties of reconstructed secondary and tertiary decay vertices.
Jets are classified as $b$-tagged if their DL1r discriminant exceeds the threshold corresponding to a $77\%$ \bHadron\ identification efficiency, at which the rejection factors for \cHadron{}s and light hadrons (from gluons, up, down, and strange quarks) are approximately 5 and 112, respectively.
Selected $b$-tagged jets are required to satisfy $\pT>\SI{20}{GeV}$ and $|\eta|<2.5$.

The missing transverse momentum \pTmiss is calculated as the negative vector sum of the $\vec{\pT}$ of all selected and calibrated objects in the event, including a term to account for energy from soft particles in the event that are not associated with any of the selected objects.
This soft term is calculated from inner-detector tracks matched to the selected primary vertex to make it more resilient to contamination from pile-up interactions~\cite{PERF-2016-07}.

Candidate \smallR\ jets with relaxed kinematic requirements of $\pT>\SI{20}{GeV}$ and $|\eta|<4.5$ before overlap removal are used as input for jet reclustering~\cite{Nachman:2014kla}.
Reclustering is performed with an \antiKt\ variable-radius algorithm~\cite{Krohn:2009zg} with $\REff=\min(1.5, \SI{550}{GeV}/\pT)$ using the transverse momentum of the resulting jet, and is found to more accurately capture the kinematics of hadronically decaying boosted heavy particles, such as top quarks and Higgs bosons, compared with fixed-radius jets with $R=1.0$, particularly at $\pT<\SI{600}{GeV}$.
Contributions from pile-up interactions and soft radiation are suppressed by removing \smallR\ subjets within each \varR\ jet that carry less than $5\%$ of its transverse momentum~\cite{Krohn:2009th}.
The resulting \varR\ jets are required to satisfy $\pT>\SI{200}{GeV}$ and $|\eta|<2.0$.

Hadronically decaying, boosted heavy-particle candidates are identified and mass-tagged as \varR\ jets with $\pT>\SI{300}{GeV}$ and mass~$m$ larger than \SI{70}{GeV}.
These mass-tagged jets are further classified as originating from vector bosons, Higgs bosons, or top quarks using analysis-specific criteria optimised for signal acceptance and mistag rate.
Vector-boson candidates ($V$-tagged jets) are mass-tagged jets with $\SI{70}{GeV}<m \leq \SI{105}{GeV}$, $\pT>\SI{350}{GeV}$, and exactly two \smallR\ subjets (\nConst{}) for $\pT<\SI{450}{GeV}$ or up to two subjets for higher \pT.
Higgs-boson candidates ($H$-tagged jets) are mass-tagged jets with $\SI{105}{GeV}<m \leq \SI{140}{GeV}$, $\pT>\SI{350}{GeV}$, and $\nConst{}=2$ for $\pT<\SI{600}{GeV}$ or $\nConst{}\leq2$ for higher \pT.
Top-quark candidates ($t$-tagged jets) are mass-tagged jets with $m>\SI{140}{GeV}$, $\pT>\SI{400}{GeV}$, and $\nConst{}\geq2$ for $\pT<\SI{700}{GeV}$ or $\nConst{}\geq1$ for higher \pT.
The selection criteria and the tagging performance are summarised in Table~\ref{tab:jet_tags}.
The performance is characterised by the true-positive rate, defined as the rate of correctly tagging a jet originating from a vector boson, Higgs boson, or top quark, and the false-positive rate, defined as the rate of incorrectly tagging a jet not originating from the corresponding particle, evaluated in \TT, \ttbar, and non-\ttbar SM events as appropriate.

\begin{table}
\centering
\caption{Selection criteria for mass-, $V$-, $H$-, and $t$-tagged jets, and their respective performances. $N_{\mathrm{const}}$ denotes the number
of \smallR\ jets retained as constituents of the reclustered variable-$R$ jet after trimming. Dashes (--) refer to definitions that are not applicable to certain categories.}
\label{tab:jet_tags}
\begin{tabular}{cccccc}
\toprule
\textbf{Jet tag} & \multicolumn{3}{c}{\textbf{Selection}} & \multicolumn{2}{c}{\textbf{Performance}}\\
\cmidrule(lr){2-4}\cmidrule(lr){5-6}
& \pT[GeV] & $m$ [GeV] & \nConst & True-positive rate & False-positive rate\\
\midrule
mass & $>300$ & $>70$ & -- & -- & --\\
$V$ & $>350$ & (70, 105] & \makecell{$=2$ if $\pT<\SI{450}{GeV}$ \\ $\leq2$ if $\pT\geq\SI{450}{GeV}$} & 0.80 & 0.20\\ %
$H$ & $>350$ & (105, 140] & \makecell{$=2$ if $\pT<\SI{600}{GeV}$ \\ $\leq2$ if $\pT\geq\SI{600}{GeV}$} & 0.65 & 0.10\\ %
$t$ & $>400$ & $>140$ & \makecell{$\geq2$ if $\pT<\SI{700}{GeV}$ \\ $\geq1$ if $\pT\geq\SI{700}{GeV}$} & 0.70 & 0.25\\ %
\bottomrule
\end{tabular}
\end{table}

Leptonic $W$-boson candidates (\Wlreco) are constructed from the lepton and the \pTmiss\ in the event:
the azimuthal direction and \pT\ of the neutrino attributed to the $W$ boson are chosen to be the same as the azimuthal direction and magnitude of \pTmiss.
The neutrino pseudorapidity $\eta^\nu$ is determined by imposing a $W$-boson mass constraint on the lepton--neutrino system.
If multiple solutions exist, the lowest~$\left|\eta^\nu\right|$ is chosen.
If no solution exists, a unique solution is constructed by setting $\eta^\nu$ to be identical with the lepton pseudorapidity and the neutrino \pT\ such that the $W$-boson mass constraint is met.

Leptonic top-quark candidates (\tlreco) are constructed from \Wlreco\ and a \smallR\ jet geometrically matched to one originating from a \bHadron.
First, the $b$-tagged jet closest in $\Delta R$ to the \Wlreco\ is identified.
The \smallR\ jet closest in $\Delta R$ to this $b$-tagged jet is then combined with \Wlreco\ to form \tlreco, provided that it lies within $\Delta R = 0.4$ of the $b$-tagged jet, within $\Delta R = 1.5$ of \Wlreco, and not within $\Delta R = 1.0$ of a $V$-, $H$-, or $t$-tagged jet.
The \smallR\ jet is used, instead of the $b$-tagged jet directly, to avoid double-counting four-momenta in the later construction of \TQuark\ candidates.
If no suitable \smallR\ jet is found, no \tlreco\ candidate is constructed.

\TQuark\ candidates are constructed from combinations of decay products.
The input set consists of all \varR~jets and, when available, the leptonic top-quark candidate~\tlreco.
To avoid double counting, the four-momentum of the \smallR\ jet associated with the \tlreco\ is subtracted from that of the nearest \varR\ jet.
If no \tlreco~candidate is reconstructed, the leptonic $W$-boson candidate~\Wlreco\ is used instead.
Two \TQuark\ candidates, \TvarR[1] and \TvarR[2], are then formed by pairing the available decay candidates and selecting the two exclusive combinations with the largest \DeltaR\ separation, reflecting the expected back-to-back topology of the \TQuark{} pair.
An analogous set of \TQuark\ candidates, \Tm[1] and \Tm[2], is constructed based on mass-tagged jets in place of \varR~jets.


\section{Analysis strategy}
\label{sec:analysis}

This search targets the pair production of \TQuark{}s in events where at least one \TQuark{} decays via $T\to Ht$, with subsequent $H\to\bbbar$ and a leptonic top-quark decay.
These signal topologies are characterised by multiple \smallR, $b$-tagged, and $H$-tagged jets, as well as moderate \met from neutrinos.
The events frequently contain an additional $t$-tagged jet, corresponding to a hadronically decaying boosted top quark, or a reconstructed leptonic top-quark candidate. %
$V$-tagged jets can arise from hadronically decaying $Z$ or $W$ bosons produced in decays of \TQuark{}s ($T\to Zt$, $T\to Wb$), Higgs bosons or top quarks.

Events are first required to satisfy a preselection targeting the general features of the signal topology, as described in Section~\ref{sec:preselection}.
Selected background processes are then corrected using data-driven reweighting to improve their modelling (Section~\ref{sec:reweighting}).
A multivariate classifier (Section~\ref{sec:MVA}) is subsequently trained to discriminate signal from SM background.
Its output, together with requirements on the $b$-tagged and $H$-tagged jet multiplicities, is used to categorise events into signal and control regions, as described in Section~\ref{sec:eventCategorisation}.
The event yields in these regions as a function of the effective mass are used in the statistical analysis.

\subsection{Preselection}
\label{sec:preselection}

A preselection is applied requiring events to satisfy the trigger selection and have at least one primary vertex candidate.
Exactly one selected electron or muon is required, suppressing multijet background, while retaining signal events in which one $W$ boson decays leptonically.
For events satisfying a single-lepton trigger, the selected lepton must match the trigger-level object within $\DeltaR < 0.15$.
Events failing to meet the single-lepton trigger acceptance or matching conditions are retained if they satisfy the \met trigger and have an offline-reconstructed \met above \SI{250}{GeV}, ensuring full trigger efficiency.
At least five \smallR\ jets ($\nJets\geq5$) and at least two $b$-tagged jets ($\nBJets\geq2$) are required, reducing contributions from non-\ttbar\ processes.
Additional conditions suppress multijet background.
The missing transverse momentum is required to satisfy $\met>\SI{20}{GeV}$ and $\mTW+\met>\SI{60}{GeV}$, where $\mTW:=\sqrt{2 p^\ell_{\mathrm T} \met (1-\cos\Delta\phi)}$ is the transverse mass of the lepton and \pTmiss\ system.
Finally, to select events consistent with the decay of high-mass particles into many objects with large transverse momentum, the effective mass \mEff, defined as the scalar sum of the \pT\ of all selected \smallR\ jets and the lepton as well as \MET, is required to exceed \SI{1000}{GeV}.

\subsection{Data-driven background reweighting}
\label{sec:reweighting}

The background samples as described in Section~\ref{sec:bkgdSamples} overestimate the multijet production cross-section, and a reweighting procedure is applied to correct the normalisation.
Reweighting factors are derived in a multijet-enriched region defined by inverting preselection requirements on \met and $\met+\mTW$, and requiring the lepton \pT\ to be below \SI{100}{GeV}, a selection under which the lepton candidate is predominantly non-prompt or misidentified.
The resulting reweighting factors of approximately 0.55 are applied to normalise all multijet distributions in the following, while the shape is retained from the MC simulation.

The background samples furthermore overestimate the cross-section of \ttbar processes at large \pT\ and underestimate it at large \nJets~\cite{TOPQ-2018-15,TOPQ-2018-17}.
Similarly, large uncertainties are observed for the modelling of \Vjets\ processes at large jet multiplicities~\cite{PMGR-2021-01}.
This is addressed with an iterative data-driven reweighting, similar to the technique used in Ref.~\cite{EXOT-2018-52}, as outlined in the following.

First, a region enriched in $Z(\to\ell\ell)$+jets events is defined to derive reweighting factors for \Vjets\ processes.
The region is designed to be kinematically close to the preselection, requiring $\nJets\geq5$, $\nBJets\geq1$, exactly two same-flavour, oppositely charged leptons, and a dilepton invariant mass consistent with the $Z$~boson.
Contamination from \ttbar processes is reduced by requiring $\met<\SI{100}{GeV}$.
Reweighting factors are then derived as a function of jet multiplicity; each factor is defined as the ratio of the difference between data and the predicted non-\Vjets\ yields to predicted \Vjets\ yields.
They range from 1.0 at $\nJets=5$ to 0.7 at $\nJets\geq8$, and are applied to all \Vjets\ processes according to their jet multiplicity.

Subsequently, a region enriched in \ttbar events is defined to derive reweighting factors for \ttbar and $tW$ processes.
The region is designed to be kinematically close to the preselection, requiring $\nJets\geq3$, $\nBJets=2$, and exactly one lepton.
Reweighting factors are first derived as a function of jet multiplicity, ranging from 0.9 at $\nJets=5$ to 1.4 at $\nJets\geq12$, and are applied to all \ttbar and $tW$ processes according to their jet multiplicity in the following.
In a final step, reweighting factors are derived as a function of $\mEffRed:=\mEff-(\nJets-5)\cdot\SI{40}{GeV}$, an observable in which \mEff\ is reduced by the average transverse momentum (\ie\SI{40}{GeV}) of each additional \smallR\ jet in \ttbar events.
This renders \mEffRed\ less dependent on \nJets\ than \mEff, though a residual dependence remains.
To account for this, reweighting factors are derived separately for four jet multiplicities ($\nJets=5,=6,=7,\geq 8$), where the binning is chosen to ensure sufficient statistical precision in the highest multiplicity bin.
This motivates the successive rather than simultaneous derivation of the \nJets\ and \mEffRed~reweighting factors.
The \mEffRed~reweighting factors are parameterised using sigmoid functions with the normalisation fixed to that determined in the preceding \nJets\ step.
The resulting factors vary between 1.3 and 0.6 as a function of \mEffRed, with a mild increase with \nJets, and are applied to all \ttbar and $tW$ processes according to both observables in the following.

\begin{figure}
\centering
\subfloat[]{%
\includegraphics[width=0.49\textwidth]{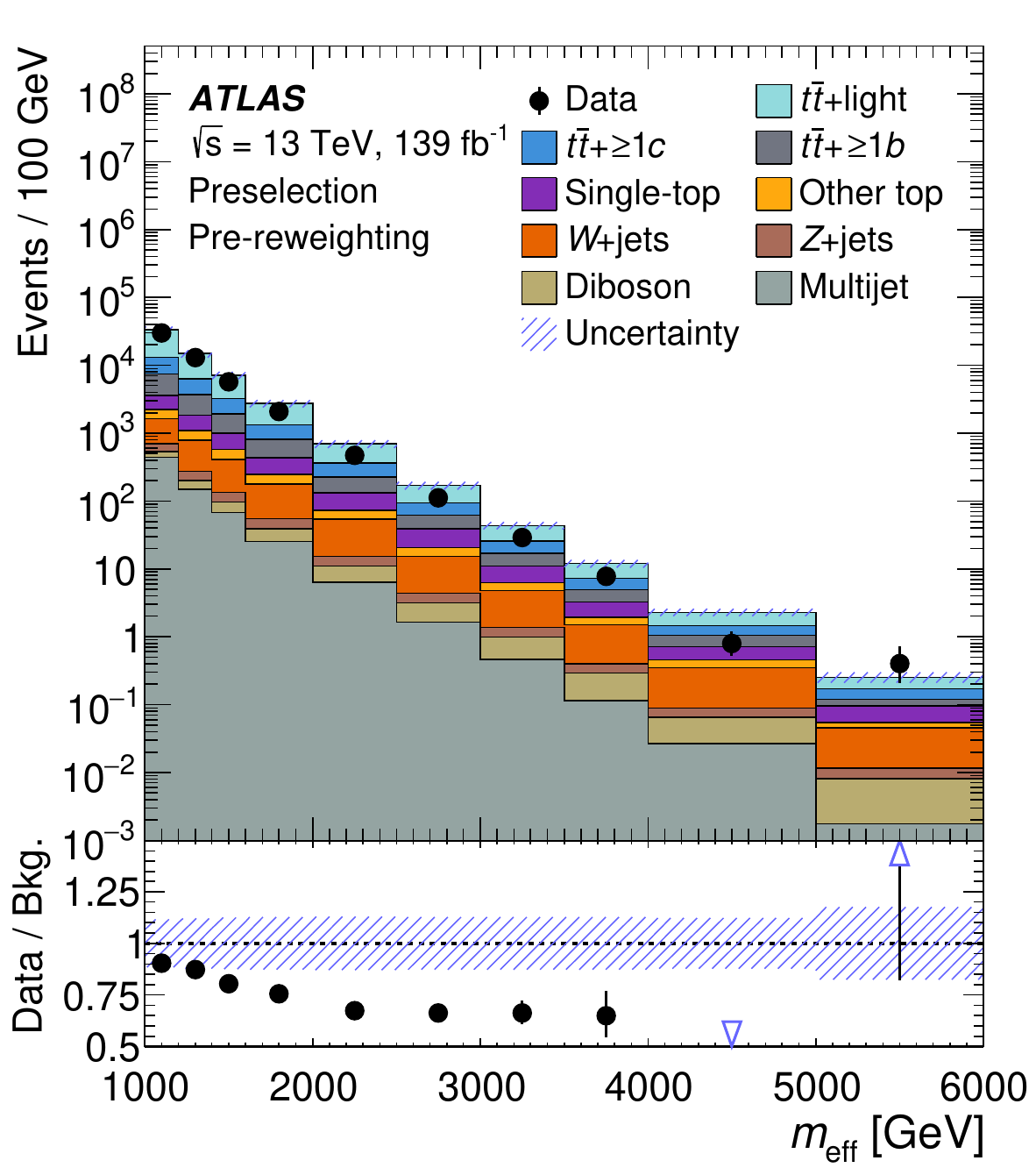}%
}
\subfloat[]{%
\includegraphics[width=0.49\textwidth]{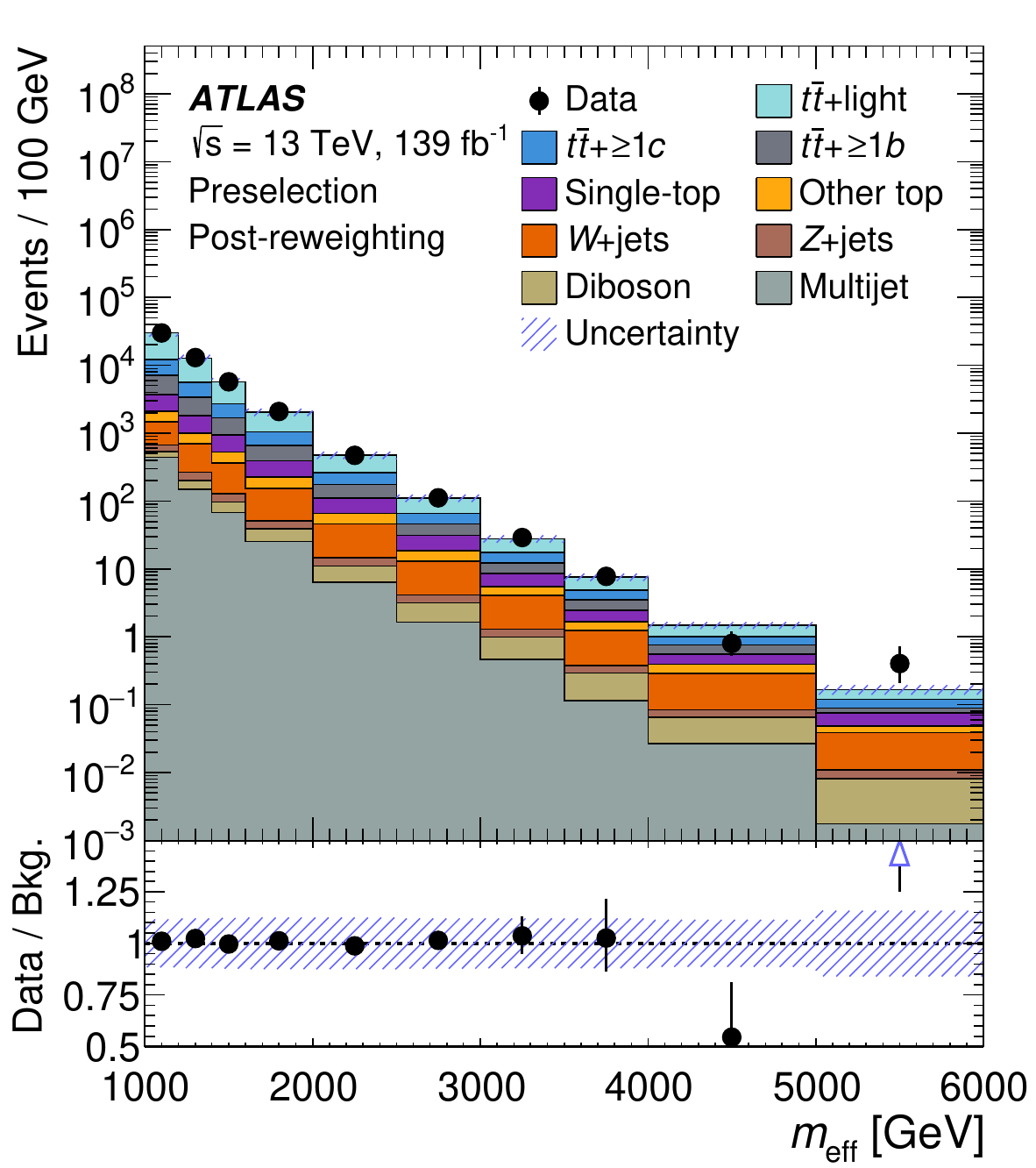}%
}
\caption{%
Distribution of the pre-fit expected yields for different SM processes and the observed data yields (black dots) after the preselection as a function of \mEff\ before (a) and after (b) the full reweighting procedure.
The bottom panel shows the ratio of the observed data yield to the total SM yield.
The blue hatched band indicates the pre-fit uncertainty in the background prediction including the statistical uncertainty, the theoretical cross-section uncertainties and the uncertainty in the integrated luminosity.
The blue triangles indicate points that are outside the vertical range of the lower pad.
}
\label{fig:kinRW}
\end{figure}

The expected and observed yields before and after the full reweighting procedure in the preselection phase space are given in Figure~\ref{fig:kinRW} as a function of \mEff, showing a significant improvement of the agreement between data and SM prediction after the reweighting.
A similar improvement is observed in \MET and the \pT and multiplicity of the other objects considered in the search such as the lepton, the leptonic $W$-boson and top-quark candidates, and the various jet collections.

\subsection{Multivariate signal classifier}
\label{sec:MVA}
Signal events are discriminated from SM backgrounds using a feed-forward neural network~(\NN) implemented as a multilayer perceptron.
The \NN\ uses 30 input variables with low mutual correlation, encompassing both kinematic and topological event properties, chosen to maximise signal-to-background discrimination.
The full list of input variables is given in Appendix~\ref{sec:appNNInputVariables}.
The \NN\ consists of four hidden layers with 30, 10, 10, and 5 nodes, respectively, and uses a sigmoid activation function.
All simulated samples for signal and background processes described in Section~\ref{sec:dataMC} are used in the training, which is performed using a binary cross-entropy loss with five-fold cross-validation to mitigate statistical bias.
Training regions are defined by requiring $\nJets\geq6$, $\nBJets\geq3$, at least three \varR\ jets ($\nVarRJets\geq3$), and at least two mass-tagged jets ($\nMJets\geq2$), in addition to the preselection.
The \NN~hyperparameters, including the architecture and activation function, are optimised by scanning multiple configurations and selecting the one that maximises the area under the receiver operating characteristic (ROC) curve in test data.

\begin{figure}[tbp]
\centering
\subfloat[]{%
\includegraphics[width=0.49\textwidth]{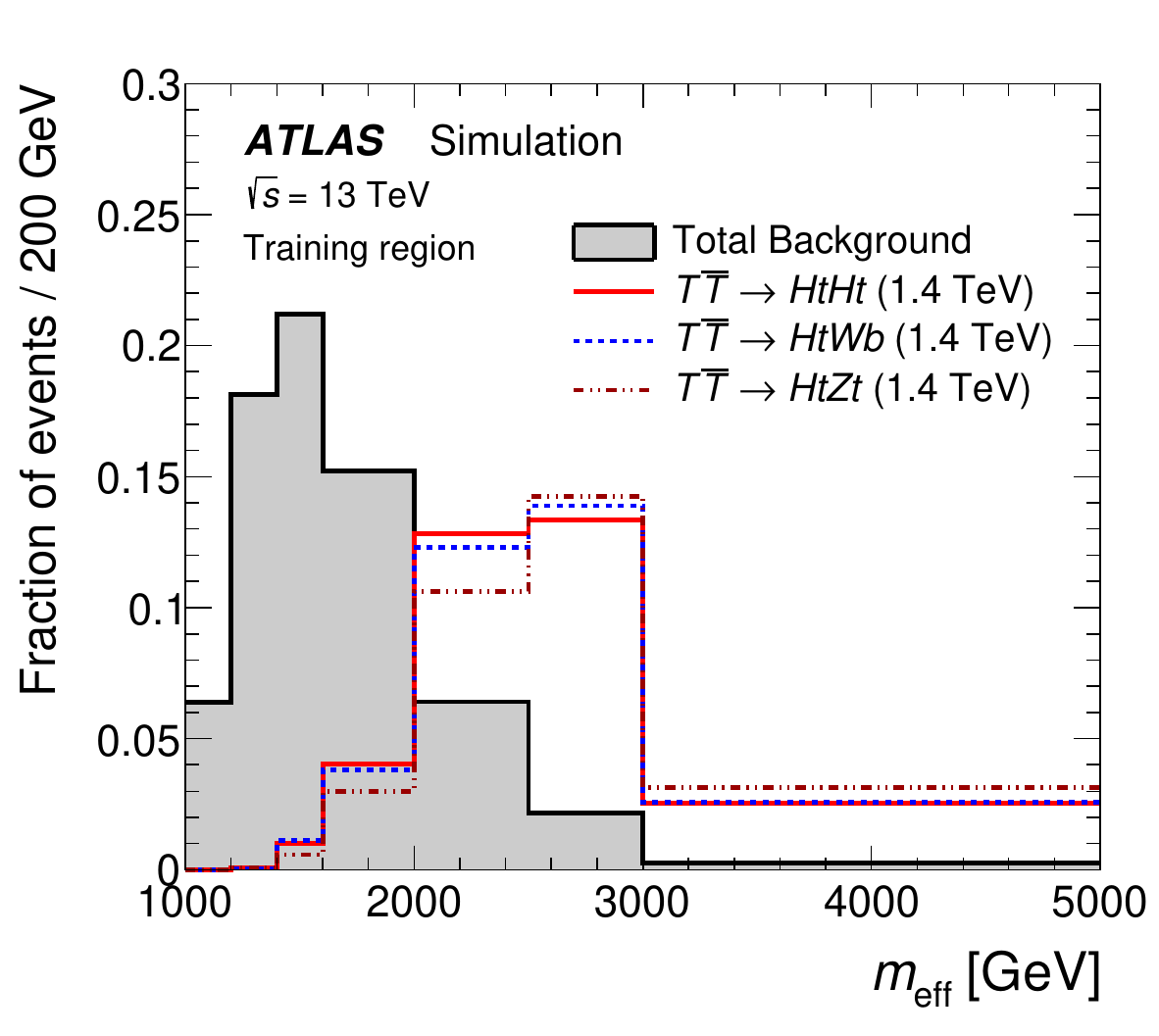}%
}
\subfloat[]{%
\includegraphics[width=0.49\textwidth]{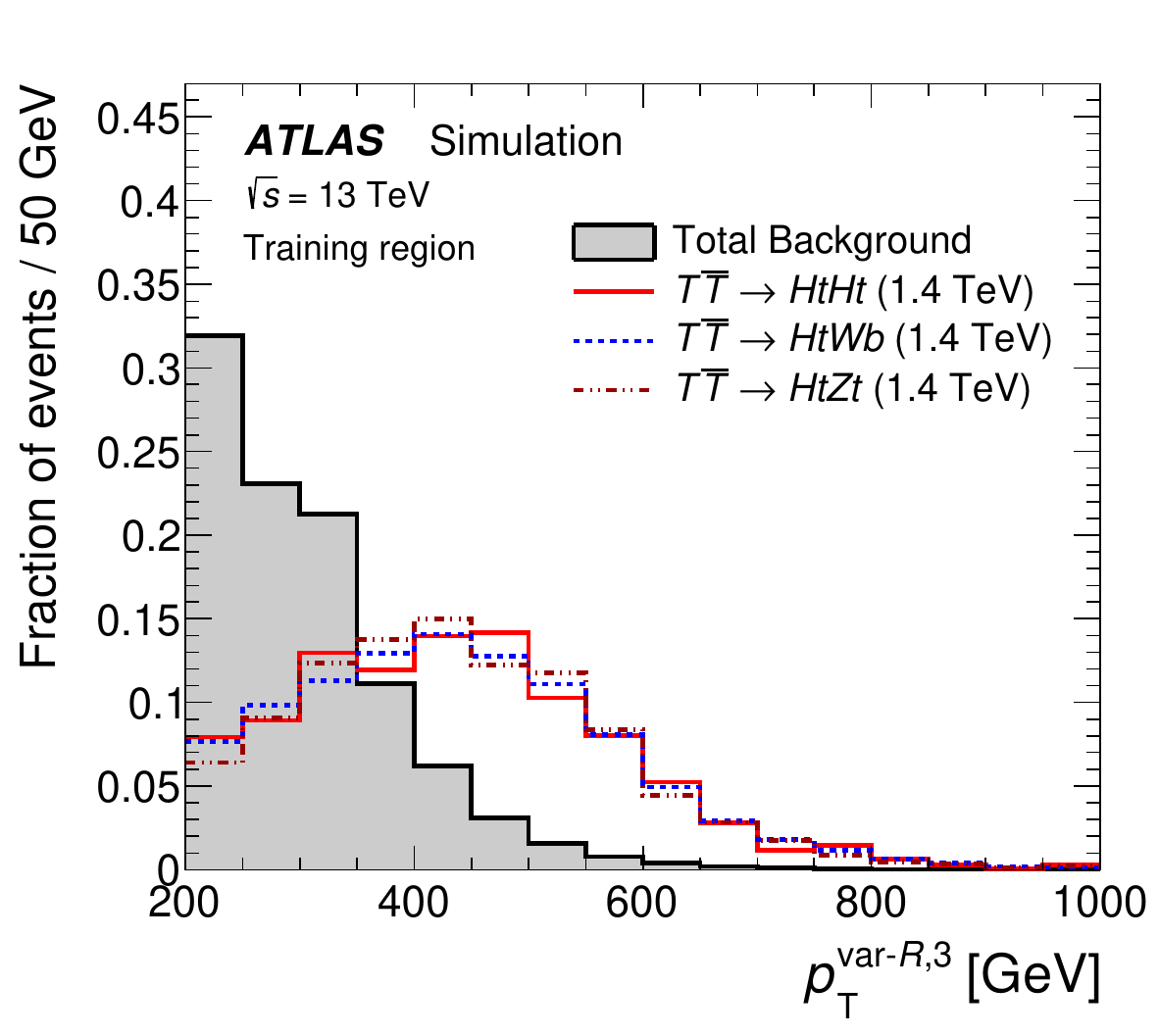}%
}
\caption{%
Normalised distributions for the total SM background (grey area) and representative signal processes (coloured lines) in the training region as a function of (a) effective mass and (b) \pT of the third-most-energetic \varR\ jet, which are the two \NN\ input variables with the largest signal-to-background separation power.
}
\label{fig:MVA_input_variables_sigBkg}
\end{figure}

\begin{figure}
\centering
\includegraphics[width=0.49\textwidth]{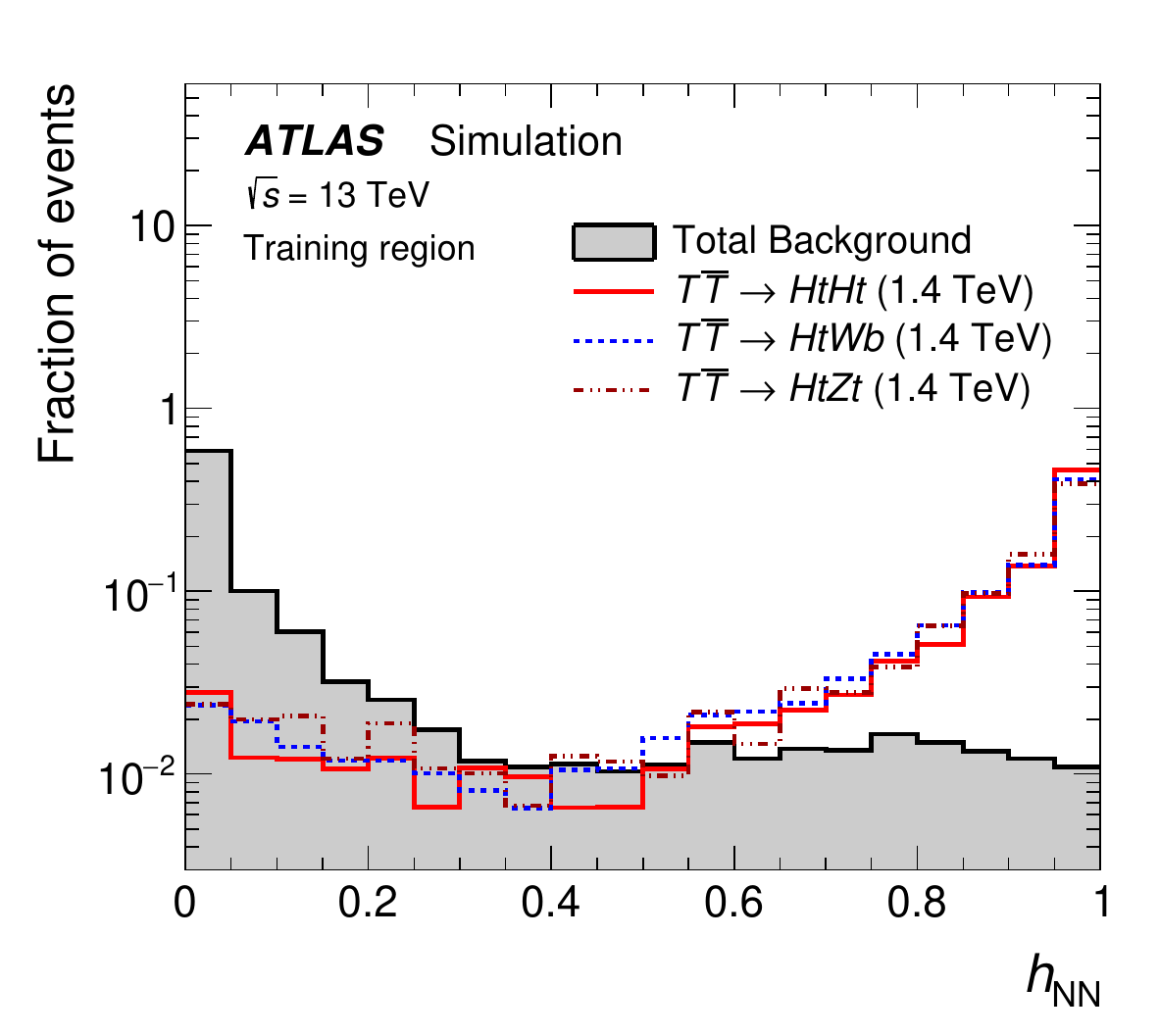}%
\caption{%
Normalised distributions for the total SM background (grey area) and representative signal processes (coloured lines) in the training region as a function of the \NN\ score \hNN.
}
\label{fig:MVA_score_sigBkg}
\end{figure}

Figure~\ref{fig:MVA_input_variables_sigBkg} gives a comparison of the yields expected from SM and representative signal processes in the training region for the two \NN\ input variables with the largest signal-to-background separation power, the effective mass~\mEff\ and the transverse momentum of the third-most-energetic \varR\ jet, \objPt[3]{\varRShort}.
The next most important input variables are the transverse momentum of the leptonic top-quark candidate, the multiplicity of \varR\ jets, and the transverse momentum of the most energetic $H$-tagged jet, as shown in Appendix~\ref{sec:appNNInputVariables}. %
The \NN\ score (\hNN) distributions for the different \TT\ decay modes have similar shapes, as shown in Figure~\ref{fig:MVA_score_sigBkg}, indicating comparable discrimination power across the signal topologies considered.
The \NN\ performance is validated by verifying the agreement between data and SM predictions in both the preselection and training region for the \NN\ input variables and output score.

\subsection{Event categorisation}
\label{sec:eventCategorisation}

For the statistical analysis, twelve signal regions~(SRs) targeting the signal topology and one control region~(CR) for estimating the \ttlight\ background are constructed.
The background modelling is further validated in two validation regions (VRs).
These three region categories (SRs, CRs, and VRs) are collectively referred to as ``analysis regions''.
Their selection requirements in addition to the preselection criteria are summarised in Table~\ref{tab:fit_regions}.

The SRs exploit the \NN\ and therefore follow the selection used for its training, as described in Section~\ref{sec:MVA}.
To improve signal sensitivity, the jet multiplicity requirement is relaxed to the preselection level of
$\nJets\geq5$, with negligible impact on the \NN\ performance.
The signal regions are subdivided according to the \NN\ score~\hNN\ into three categories of decreasing signal purity.
The high-\NN~(\HMVA) region, defined by $\hNN>0.81$, achieves a background rejection factor above 20 for an \HtHt\ signal efficiency of $79\%$.
The mid-\NN~(\MMVA) region, defined by $\hNNLow\leq\hNN\leq\hNNHigh$, retains $17\%$ of \HtHt\ signal events.
The low-\NN~(\LMVA) region, defined by $\hNN<\hNNLow$, has a background rejection factor of 4 while accepting only $4\%$ of \HtHt\ signal events, resulting in a large fraction of \ttcin\ and \ttbin\ events that makes it well-suited for constraining these backgrounds in the fit.
Since the \NN\ is trained to be agnostic to the vector-like quark decay mode, sensitivity to the \TTtoHtHt\ signal is enhanced by further subdividing the signal regions according to the $b$-tagged jet multiplicity ($3$ or $\geq4$) and, independently, whether at least one $H$-tagged jet is present.

For the \ttlight\ control region, events are required to satisfy the preselection, to contain exactly one reconstructed leptonic top-quark candidate \tlreco\ ($\nTlreco=1$), and to have exactly one $V$-tagged or $t$-tagged jet ($\nJets[V]+\nJets[t]=1$).
Events containing $H$-tagged jets are vetoed ($\nJets[H]=0$).
A high purity in \ttlight\ events is achieved by requiring exactly two $b$-tagged jets.
Two validation regions are defined to validate the modelling of the \ttcin\ and \ttbin\ backgrounds in the SRs.
They follow the same selections as the control region, but require exactly three or at least four $b$-tagged jets, respectively.
To increase statistical precision, no requirement on \tlreco\ is imposed.
The validation regions are not orthogonal to the $0H$ signal regions and are therefore not used in the simultaneous fit.

\FloatBarrier

\begin{table}
\centering
\caption{Summary of requirements for control, validation and signal regions in addition to the preselection requirements. Dashes (--) refer to selection requirements that are not applied.}
\label{tab:fit_regions}
\resizebox{\textwidth}{!}{%
\begin{tabular}{cccccccccccccccc}
\toprule
\textbf{Observable} & \multicolumn{15}{c}{\textbf{Selection}} \\
\cmidrule(lr){2-2}\cmidrule(lr){3-4}\cmidrule(lr){5-16}
\multirow{3}{*}{\color{lightgray}\makecell[r]{region\\ name} $\left\{\rule{0pt}{22pt}\right.$\hspace{-35pt}} & CR & \multicolumn{2}{c}{VRs} & \multicolumn{4}{c}{\LMVA\ SRs} & \multicolumn{4}{c}{\MMVA\ SRs} & \multicolumn{4}{c}{\HMVA\ SRs} \\
& $2b$ & $3b$ & $4b$ & \multicolumn{2}{c}{$3b$} & \multicolumn{2}{c}{$4b$} & \multicolumn{2}{c}{$3b$} & \multicolumn{2}{c}{$4b$} & \multicolumn{2}{c}{$3b$} & \multicolumn{2}{c}{$4b$} \\
&&&& $0H$ & $1H$ & $0H$ & $1H$ & $0H$ & $1H$ & $0H$ & $1H$ & $0H$ & $1H$ & $0H$ & $1H$\\
\cmidrule(lr){1-1}\cmidrule(lr){2-4}\cmidrule(lr){5-16}
\nJets[H] & \multicolumn{3}{c}{$=0$} & $=0$ & $\geq1$ & $=0$ & $\geq1$ & $=0$ & $\geq1$ & $=0$ & $\geq1$ & $=0$ & $\geq1$ & $=0$ & $\geq1$ \\
$\nJets[V]+\nJets[t]$ & \multicolumn{3}{c}{$=1$} & \lightDash & \lightDash & \lightDash & \lightDash & \lightDash & \lightDash & \lightDash & \lightDash & \lightDash & \lightDash & \lightDash & \lightDash \\
\nBJets & $=2$ & $=3$ & $\geq4$ & \multicolumn{2}{c}{$=3$} & \multicolumn{2}{c}{$\geq4$} & \multicolumn{2}{c}{$=3$} & \multicolumn{2}{c}{$\geq4$} & \multicolumn{2}{c}{$=3$} & \multicolumn{2}{c}{$\geq4$} \\
\nTlreco & $=1$ & \lightDash & \lightDash & \lightDash & \lightDash & \lightDash & \lightDash & \lightDash & \lightDash & \lightDash & \lightDash & \lightDash & \lightDash & \lightDash & \lightDash \\
\nVarRJets & \lightDash & \lightDash & \lightDash & \multicolumn{12}{c}{$\geq3$} \\
\nMJets & \lightDash & \lightDash & \lightDash & \multicolumn{12}{c}{$\geq2$} \\
\hNN & \lightDash & \lightDash & \lightDash & \multicolumn{4}{c}{$<\hNNLow$} & \multicolumn{4}{c}{$[\hNNLow, \hNNHigh]$} & \multicolumn{4}{c}{$>\hNNHigh$} \\
\bottomrule
\end{tabular}
}
\end{table}



\section{Systematic uncertainties}
\label{sec:uncertainties}

Systematic uncertainties can be categorised into two main types: experimental uncertainties and modelling uncertainties.
Experimental uncertainties arise from the detector response to the various physics objects.
Modelling uncertainties encompass uncertainties related to theoretical predictions and the parameters chosen for those predictions, and to the data-driven corrections.
The former arise from missing higher-order corrections, parton-shower modelling, alternative PDFs, and assumptions in the underlying theory models.
Each systematic uncertainty is considered to be correlated across processes, channels, and bins of the signal discriminant, unless explicitly stated otherwise.
Uncertainties from different sources are considered to be uncorrelated with each other.

The leading sources of uncertainty in this search arise from the modelling of the $t\bar{t}$ and single-top $Wt$
backgrounds, the flavour-tagging efficiencies ($b$, $c$, and light), and the jet mass resolution.
The relative contributions of these uncertainties vary across analysis regions, and since different signal processes populate the analysis regions in different proportions, their impact on the search sensitivity is signal-process dependent.
The considered systematic uncertainties are detailed in the following.

\subsection{Experimental uncertainties}

Uncertainties associated with jets arise from the jet-energy scale (JES), jet-energy resolution (JER), the jet-mass
scale (JMS), jet-mass resolution (JMR), and the efficiency of the JVT requirements. The JES and JER uncertainties
are estimated by combining information from collision data,
test-beam data, and simulation~\cite{JETM-2018-05, JETM-2018-02}.
The JES (JER) uncertainties are split into 30 (8) uncorrelated components, corresponding to different physical sources.
The uncertainties in the JMS are measured using the ratio of calorimeter and track measurements of the jet mass, and are split into 8 uncorrelated components.
The uncertainty in the JMR is estimated by comparing each nominal sample
with an alternative event sample in which the mass of each jet is smeared by a Gaussian function whose
width is increased by $20\%$ relative to the nominal JMR.
These jet-related uncertainties in \smallR\ jets are propagated to the \varR\ jets in the reclustering procedure.

The flavour-tagging efficiencies for $b$-, $c$-, and light-tagged jets in simulation are corrected to match those measured in \ttbar- and \Zjets-enriched data, following the methods described in Refs.~\cite{FTAG-2018-01, FTAG-2020-08, FTAG-2019-02}.
Uncertainties are evaluated using nine independent sources for $b$-tagged jets, five for $c$-tagged jets, and six for light-flavour jets.
These sources are treated as uncorrelated between the different jet flavours.
An additional extrapolation uncertainty is included for high-\pT jets beyond the kinematic reach of the calibration data.
This uncertainty is assumed to be correlated across all jet flavours.

Uncertainties associated with leptons arise from the efficiencies of the trigger selection, reconstruction,
identification, and isolation criteria, as well as the lepton momentum scale and resolution. These are
measured in data using $Z \rightarrow \ell\ell$ and $J/\psi \rightarrow \ell\ell$ decays~\cite{PERF-2015-10, EGAM-2018-01}.

The estimated uncertainty in the integrated luminosity of the combined 2015–2018 data is $1.7\%$~\cite{ATLAS-CONF-2019-021}.
Uncertainties associated with the resolution and scale of \met~\cite{PERF-2016-07}, as well as with the efficiency of the \met trigger are also considered.
The uncertainty associated with the modelling of pile-up interactions is evaluated from variations of the pile-up reweighting applied to simulated events.

\subsection{Theoretical modelling uncertainties}
\label{sec:theory_uncertainties}

Some systematic uncertainties arise in the theoretical modelling of the signal and background
processes detailed in Section~\ref{sec:dataMC}.
All simulated SM processes include predictions for alternative choices of the factorisation and renormalisation scales, and from standard PDF-set variations that capture the statistical and modelling uncertainties in the PDF fits.

Systematic uncertainties in the signal modelling arising from the choice of PDF set are evaluated following the \PDFforLHC[15] prescription~\cite{Butterworth:2015oua}, using the \MSTW~\NLO, \CTten~\NLO~\cite{Lai:2010vv,Gao:2013xoa}, and \NNPDF[3.0nlo] PDF sets.
No further systematic uncertainties in the signal modelling, in particular regarding the production cross-section, are considered.

An uncertainty of approximately $\pm6\%$ is assigned to the inclusive \ttbar production cross-section~\cite{Czakon:2011xx} for \ttlight\ processes.
For single-top processes, a $\pm5\%$ uncertainty in the total cross-section, estimated as a weighted average of the theoretical uncertainties in $Wt$-, $s$-, and $t$-channel production~\cite{Kidonakis:2011wy, Kidonakis:2010ux, Kidonakis:2010tc}, is included.
Multiple additional sources of systematic uncertainty affecting the modelling of \ttbar and single-top production are considered.
Uncertainties due to the choice of the \NLO generator are estimated by comparing the
nominal \POWPY samples with alternative samples generated by \MGNLO and
showered by \Pythia. The nominal samples are compared with \POWHER samples to estimate
the uncertainties in the modelling of the parton showering and hadronisation processes. The alternative
samples used to evaluate modelling uncertainties are detailed in Section~\ref{sec:bkgdSamples}. The uncertainty
due to initial-state radiation is estimated by simultaneously varying the $h_{\mathrm{damp}}$ parameter and the
renormalisation and factorisation scales, and choosing the Var3c up/down variants of the A14 set of tuned parameters as described in Ref.~\cite{ATL-PHYS-PUB-2017-007}.
The impact of final-state radiation is evaluated by doubling or halving the renormalisation scale for emissions from the parton shower.
The systematic uncertainties arising from the choice of PDF set are evaluated following the \PDFforLHC[15] prescription identical to those for the signal modelling.
All \ttbar and single-top uncertainties are considered to be uncorrelated among \ttlight, \ttcin, \ttbin, and single-top samples, but correlated across the subcategories of the single-top background ($Wt$-, $s$-, or $t$-channel).
An additional uncertainty is assigned for the modelling of \ttbin\ processes by comparing the nominal five-flavour scheme \ttbar sample with the alternative \ttbb\ sample generated in the four-flavour scheme, in which \bbbar\ pairs enter the matrix element.
The magnitude of this uncertainty is negligible except in analysis regions with $\geq$4$b$, where it ranges from $9\%$ in VR~$4b$ to $67\%$ in SR~$4b$-$1H$-\HMVA, the latter reflecting the susceptibility of the \NN\ to small kinematic changes and the limited statistical precision on the estimation procedure in this region.
Despite the significant size of this uncertainty in certain low-yield bins, its impact on the exclusion limits remains negligible.

An additional systematic uncertainty in $Wt$-channel production, concerning the separation between \ttbar and $Wt$ at \NLO~\cite{Frixione:2008yi}, is assessed by comparing the nominal sample, which uses the diagram removal scheme, with the alternative sample using the diagram subtraction scheme but otherwise identical event generation settings.

The uncertainties in the modelling of the \Vjets\ backgrounds are estimated by varying the values of the
internal renormalisation and factorisation scale parameters in \Sherpa. An additional $\pm30$\% normalisation
uncertainty for the \Vjets\ background is considered for events in analysis regions with different $b$-tag
multiplicity (2, 3, $\ge4$), uncorrelated among $b$-tag multiplicities. This uncertainty is based on variations of
the factorisation and renormalisation scales and \Sherpa matching parameters~\cite{HIGG-2020-23}, and is considered to be
correlated between $W$+jets and $Z$+jets processes.

The kinematic reweighting of the main background processes (described in Section~\ref{sec:reweighting}) is derived separately for each modelling uncertainty.
This ensures that each alternative background model reproduces the data in the regions used to derive the reweighting factors.
After reweighting, the modelling uncertainties therefore primarily account for extrapolations in kinematics and background composition from the derivation regions to the analysis regions.
Additional uncertainties arise from the reweighting procedure itself.
For the multijet reweighting, a $\pm11\%$ normalisation uncertainty is assigned based on the statistical precision in the derivation region; its impact on the sensitivity is negligible due to the small multijet contribution.
For the \ttbar and \Vjets\ reweightings as a function of \nJets, the uncertainty is given by the bin-by-bin statistical uncertainties in the derivation regions.
For the \mEffRed-dependent reweighting, the uncertainty is obtained by propagating the fit parameter uncertainties according to their covariance matrix, thereby accounting for their correlations.
The functional form is kept fixed as a sigmoid function.
To obtain a conservative estimate that covers both normalisation and shape variations, a shift of two standard deviations is applied.
The resulting uncertainty ranges from $\pm0.5$ to $\pm12\%$, depending on the jet multiplicity and reweighted process, and is largest for events with $\nJets\geq8$.

Diboson, $\ttbar W$, \ttZy, $\ttbar H$, $tH$ and $tZ$ processes contribute only a small fraction of events in the analysis regions.
Their shape uncertainties are therefore neglected, and only normalisation uncertainties are considered.
These are taken from PDF and scale variations in \NLO\ calculations, resulting in $\pm15\%$ for $\ttbar W$, \ttZy, $tH$ and $tZ$, $+9/-12\%$ for $\ttbar H$, and $\pm30\%$ for the four-top production.
For the diboson background, a $\pm5\%$ uncertainty is assigned to the inclusive cross-section at \NLO~\cite{Campbell:1999ah}.
An additional uncorrelated $\pm24\%$ uncertainty per additional jet is included to account for modelling of additional radiation, based on comparisons of different \LO~matrix-element and parton-shower merging schemes~\cite{Alwall:2007fs}.
This is evaluated at the average jet multiplicity in the analysis regions, which is five jets, corresponding to three additional jets from radiation.
A further $\pm30\%$ normalisation uncertainty is assigned to the production of additional heavy-flavour jets, following the treatment used for \Vjets~\cite{HIGG-2020-23}.
The total normalisation uncertainty for diboson processes amounts to $\pm56\%$.
As the two leading $b$-tagged jets in diboson events predominantly originate from $W \rightarrow cs$ or $Z \rightarrow bb$ decays, this uncertainty is applied only in analysis regions with $\ge3$ $b$-tagged jets.
All of these uncertainties are added in quadrature within each region, and the resulting uncertainty is treated as uncorrelated between the $2b$ and $\geq 3b$ analysis regions.


\begin{table}[tbp]
\centering
\caption{%
Summary of the different sources of uncertainty in $\mu$, shown for a vector-like \TQuark{} with mass~\SI{1.4}{TeV} under the \singlet, \doublet, and \BTtoHtOne\ scenarios.
The impact of a group of nuisance parameters is obtained by summing in quadrature the impacts of all nuisance parameters in this category.
The fractional contribution of an uncertainty source gives the ratio of the uncertainty in $\mu$ resulting from the fit when fixing the nuisance parameters corresponding to each category to the uncertainty in the best-fit value of $\mu$.
}
\label{tab:results_likelihood_impact}
\begin{tabular}{l*{4}{S[table-format=2.1]}}
\toprule
& \multicolumn{3}{c}{Fractional contribution [\%], $m_T=\SI{1.4}{TeV}$} \\
\midrule
\textbf{Uncertainty source} & \textbf{\singlet} & \textbf{\doublet} & \textbf{\BTtoHtOne} \\
\midrule
Theory uncertainties (per process group) \\
\indentTable \ttbar\\
\indentTable\indentTable Modelling                    & 21   & 24   & 26   \\
\indentTable\indentTable Reweighting                  & 27   & 24   & 21   \\
\indentTable Single-top                               & 14   & 10   &  6.3 \\
\indentTable Non-top                                  &  6.2 &  4.7 &  4.0 \\
\indentTable Other top                                &  4.9 &  4.6 &  3.8 \\
Experimental uncertainties \\
\indentTable Jets                                     & 17   & 15   & 15   \\
\indentTable Flavour-tagging                          &  8.0 &  8.0 &  8.2 \\
\indentTable Leptons, pile-up, \MET \& luminosity     &  1.9 &  1.7 &  1.8 \\
\midrule
Total systematic uncertainty                          & 42   & 40   & 38   \\
\midrule
Data statistical uncertainty                          & 90   & 90   & 91   \\
MC statistical uncertainty \\
\indentTable Background                               & 14   & 14   & 14   \\
\indentTable Signal                                   &  1.4 &  3.0 &  5.3 \\
\ttbar normalisation		                          &  3.8 &  3.7 &  4.1 \\
\midrule
Total statistical uncertainty                         & 91   & 92   & 92   \\
\bottomrule
\end{tabular}
\end{table}

%
%
%
%
%
%
%
%
%
%
%
%
%
%
%
%
%
%
%
%
%
%
%
%
%
%
%
%
%
%
%
%
%
%
%
%
%


\section{Statistical analysis}
\label{sec:stat_analysis}

For each benchmark scenario considered in this search, the \mEff\ distributions across all signal and control regions are jointly studied to test for the presence of the predicted signal.
A binned likelihood function~\fullLh~\cite{Rolke:2004mj} is constructed as a product of Poisson probability terms over all \mEff\ bins considered in the search.
The likelihood function depends on the signal-strength parameter $\mu$, which multiplies the predicted production cross-section for the signal, free normalisation factors $\kappa$, and $\theta$, a set of nuisance parameters that encode the effect of systematic uncertainties in the signal and background expectations.
The normalisation factors $\kappa$ are applied separately for the \ttcin\ and \ttbin\ backgrounds, which proved beneficial in other ATLAS analyses with three or more $b$-tagged jets~\cite{HIGG-2017-03,TOPQ-2019-03,HMBS-2024-46}.
Nuisance parameters accounting for statistical uncertainties from the limited size of the simulated samples are modelled in the likelihood function using Poisson constraints.
All other nuisance parameters are modelled with Gaussian constraints.

For a given value of $\mu$, variations in the normalisation factors $\kappa$ and nuisance parameters $\theta$ allow the expectations for signal and background to change according to the corresponding systematic uncertainties.
The fitted values of $\kappa$ and $\theta$ correspond to deviations from the nominal expectations that globally provide the best fit to the data.
This procedure reduces the impact of systematic uncertainties on the search sensitivity and improves the background prediction by taking advantage of the highly populated, background-dominated control region and \LMVA\ signal regions included in the likelihood fit.

The test statistic $q_\mu$, as implemented in the \TRExFitter framework~\cite{TRExFitter} based on \RooStats~\cite{Verkerke:2003ir,Moneta:2010pm}, is defined as the profile likelihood ratio:
$q_\mu := -2\ln(\lh(\mu,\hathatKappa,\hathatTheta)/\fullLhHat)$.
Here, $\hat{\mu}$, $\hat{\kappa}$, and $\hat{\theta}$ are the values of the parameters $\mu$, $\kappa$ and $\theta$, respectively, which simultaneously maximise the likelihood function~\fullLh.
The values of the normalisation factors and nuisance parameters which maximise the likelihood function for a given value of $\mu$ are denoted by \hathatKappa and \hathatTheta\!\kern-1pt, respectively.
The statistic used for the discovery test, \ie to determine the compatibility of the observed data with the background-only hypothesis, is obtained by setting $\mu=0$ in the profile likelihood ratio and leaving $\hat{\mu}$ unconstrained:
$q_0 = -2\ln(\lh(0,\hathatKappa[0],\hathatTheta[0])/\fullLhHat)$.
The $p$-value of the discovery test is given by the integral of the probability distribution of $q_{0}$ above the observed $q_{0}$ value when assuming the background-only hypothesis, computed using the asymptotic approximation detailed in Refs.~\cite{Cowan:2010js,ErratumCowan:2010js}.
The test statistic for deriving the upper limits on the signal production cross-section for each signal scenario considered sets $0\leq \hat{\mu} \leq \mu$ and is computed using $q_\mu$ in the \CLs\ method~\cite{Junk:1999kv,Read:2002hq}, also in the asymptotic approximation.
For a given signal scenario, values of the production cross-section (parameterised by $\mu$) yielding $\CLs < 0.05$ are excluded at $\geq 95\%$ confidence level (\CL).


\section{Results}
\label{sec:results}

\subsection{Likelihood fit}
\label{sec:results_likelihoodFit}

\begin{figure}[tbp!]
\centering
\includegraphics[width=0.8\textwidth]{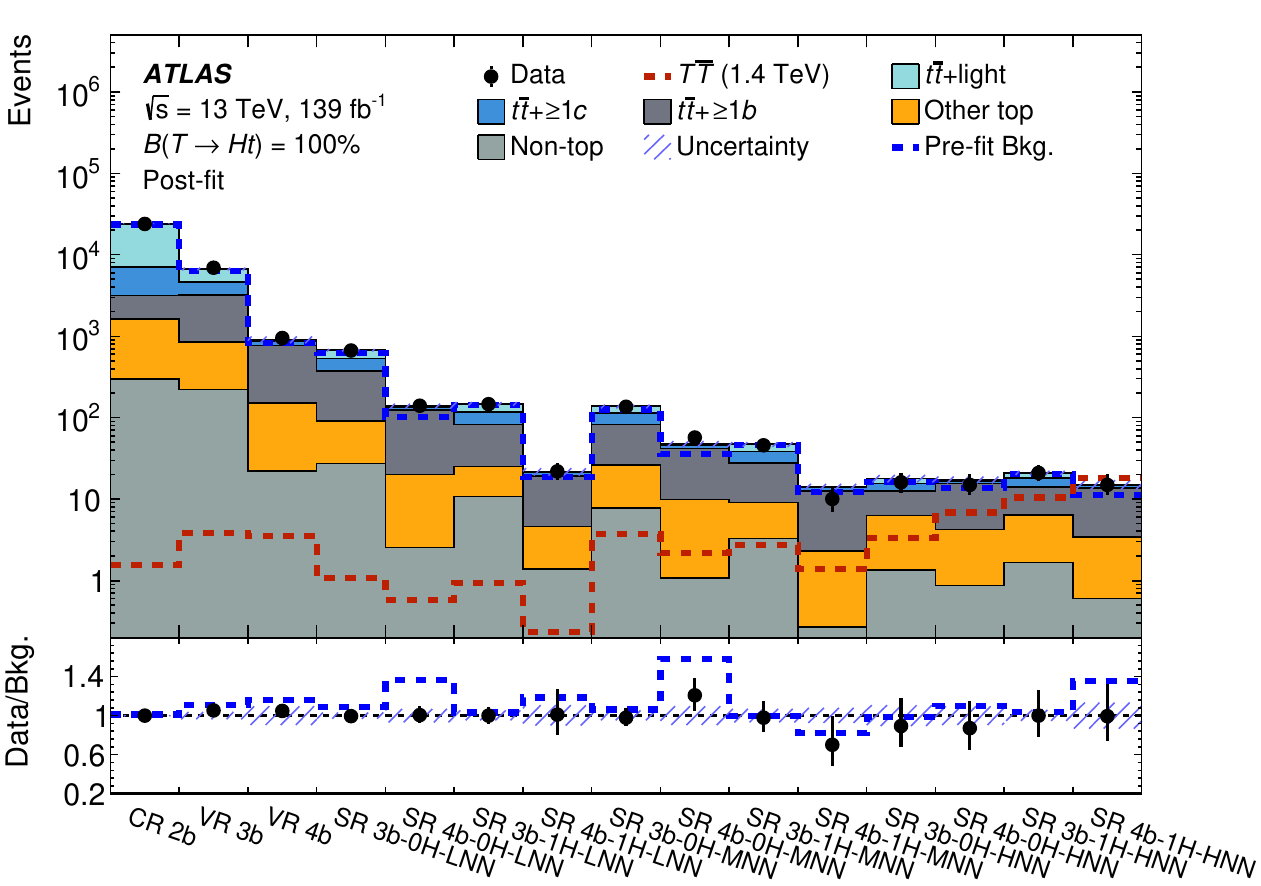}
\caption{%
Comparison between total observed yields (black dots) and expected SM yields (coloured areas) in all analysis regions after a fit under the background-only hypothesis in the control and signal regions.
The fit result is extrapolated to the validation regions, which are not included in the fit.
The bottom panel displays the ratio of data and total SM background.
The size of the combined statistical and systematic uncertainty in the background prediction is indicated by a blue hatched band, including all sources of systematic uncertainty described in Section~\ref{sec:uncertainties}.
In the upper panel, the dashed blue (red) lines give the total expected SM yield (expected yield from a \BTtoHtOne\ signal with $m_T=\SI{1.4}{TeV}$) before the fit.
In the lower panel, the dashed blue line gives the ratio of the observed yield to the total expected SM yield before the fit.
}
\label{fig:results_likelihood_summary_postfit}
\end{figure}

A likelihood fit is performed under the signal-plus-background hypothesis for each of the generated signal points.
No significant excess of the data over the SM prediction, indicative of a signal, is observed.
Figure~\ref{fig:results_likelihood_summary_postfit} shows results for a likelihood fit under the background-only hypothesis, giving the comparison between the total observed and expected yields in each analysis region after the fit.
The combined impact of the systematic uncertainties is constrained in the fit, making use of the information from the large statistics in the signal-depleted control region and \LMVA\ signal regions.
The normalisation factors for the \ttcin\ and \ttbin\ backgrounds are estimated from the fit to be $\kappa_{\ttcin}=0.97\pm0.40$ and $\kappa_{\ttbin}=1.20\pm0.23$, respectively, which is compatible with the findings of comparable ATLAS analyses~\cite{HIGG-2017-03,TOPQ-2019-03,HMBS-2024-46}.
The fit result is extrapolated to the validation regions, which are not included in the fit, to verify the reliability of the background modelling.

\begin{figure}[htbp!]
\centering
\subfloat[]{\includegraphics[width=0.49\textwidth]{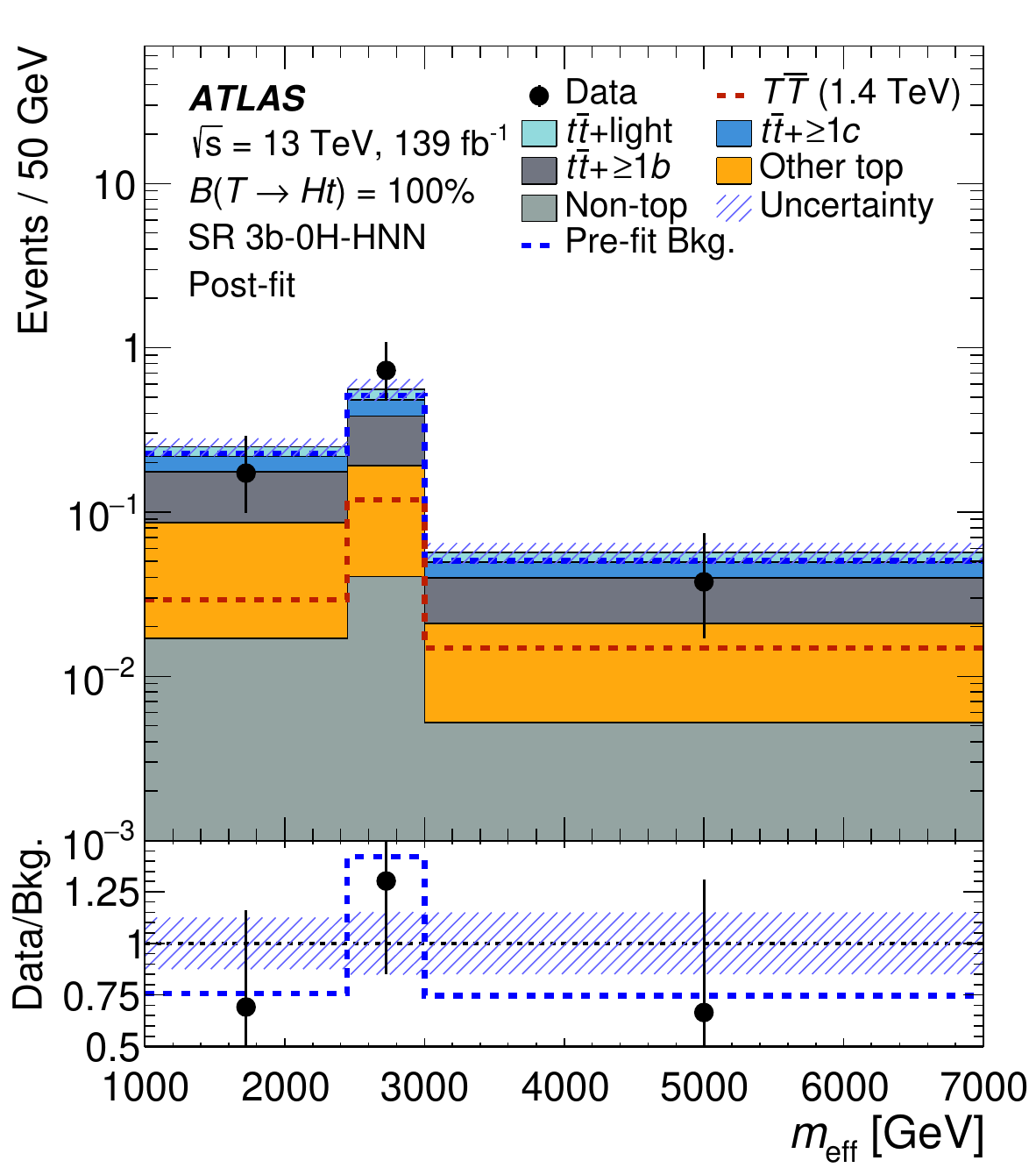}}
\subfloat[]{\includegraphics[width=0.49\textwidth]{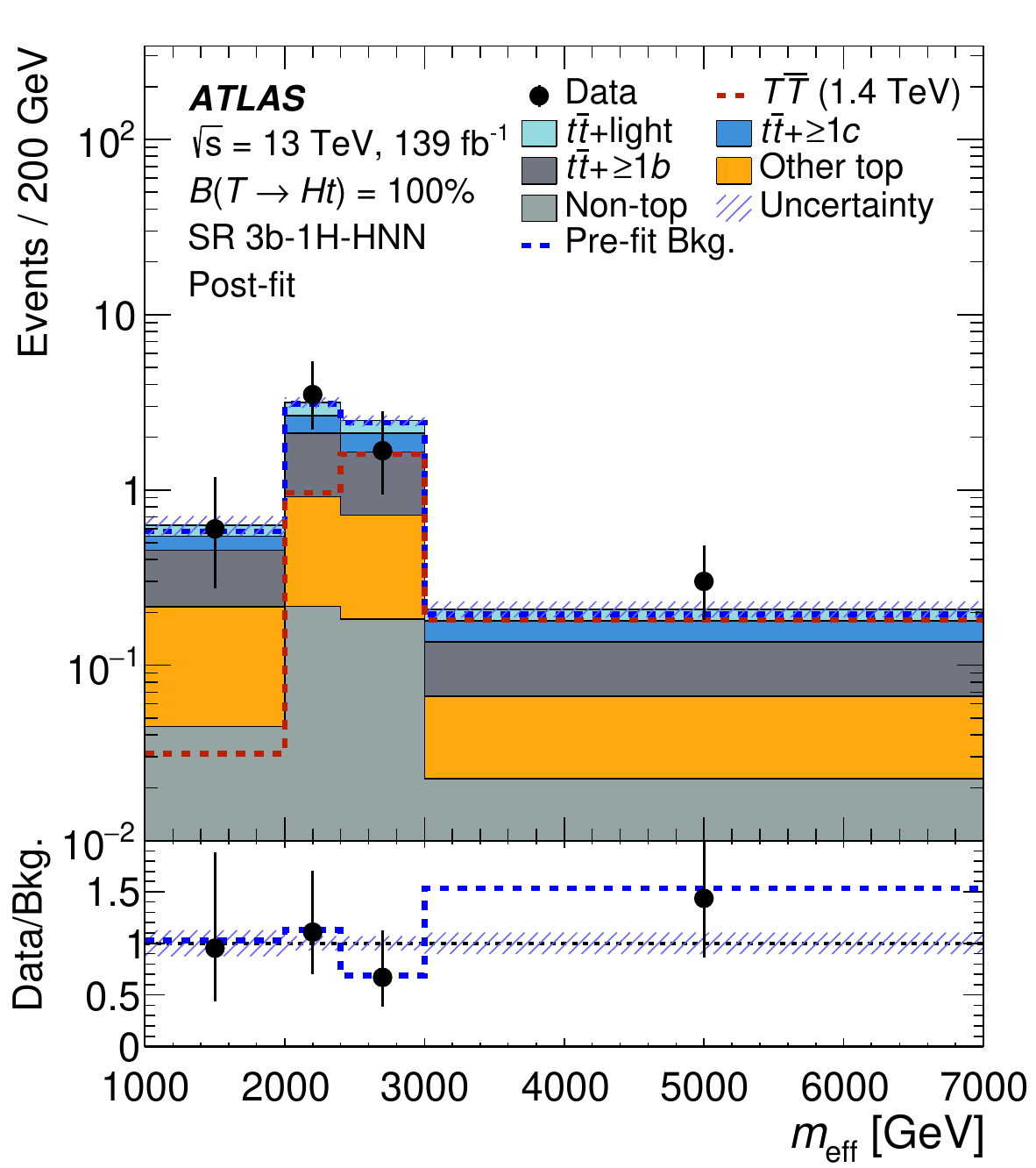}}\\
\subfloat[]{\includegraphics[width=0.49\textwidth]{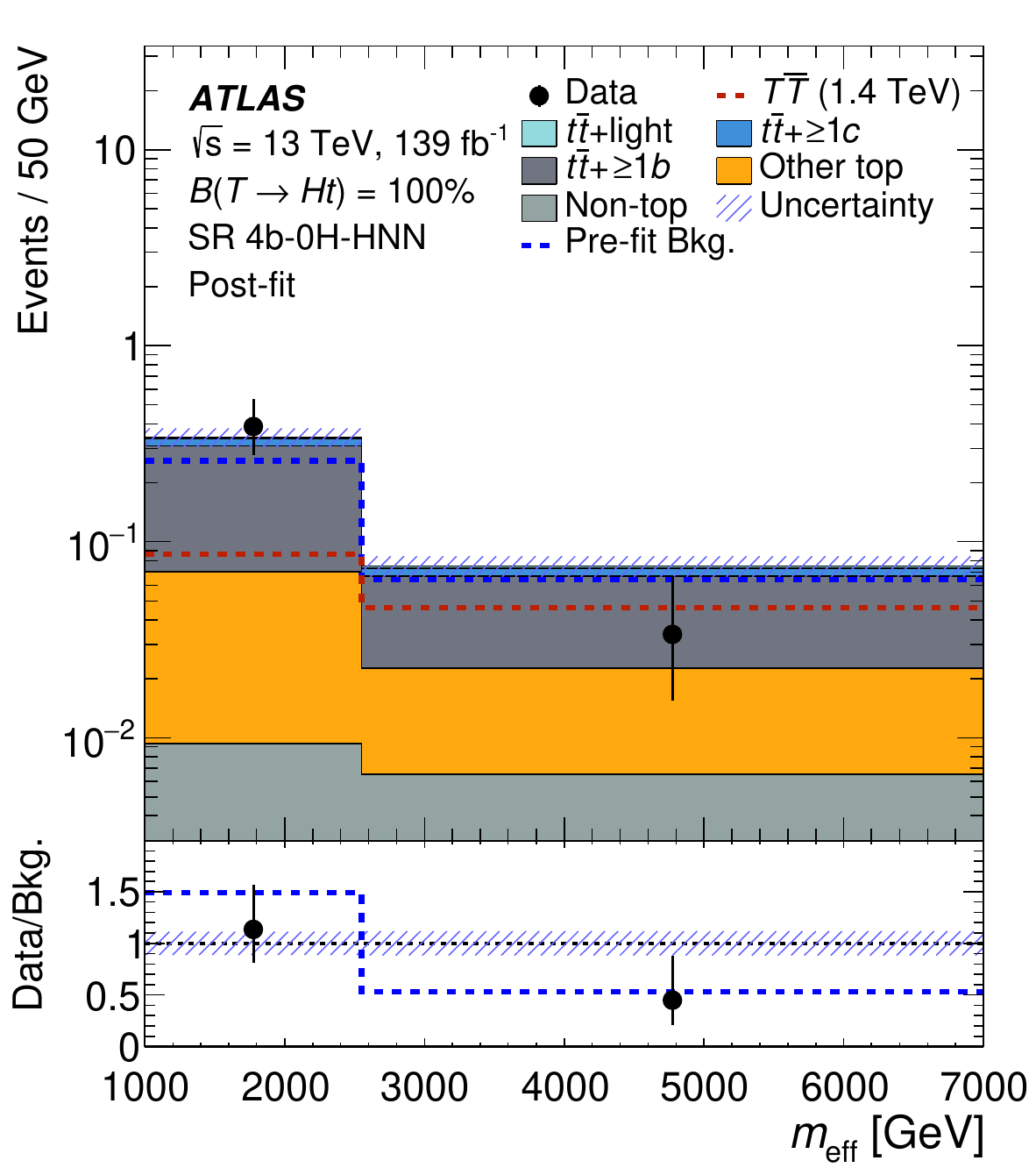}}
\subfloat[]{\includegraphics[width=0.49\textwidth]{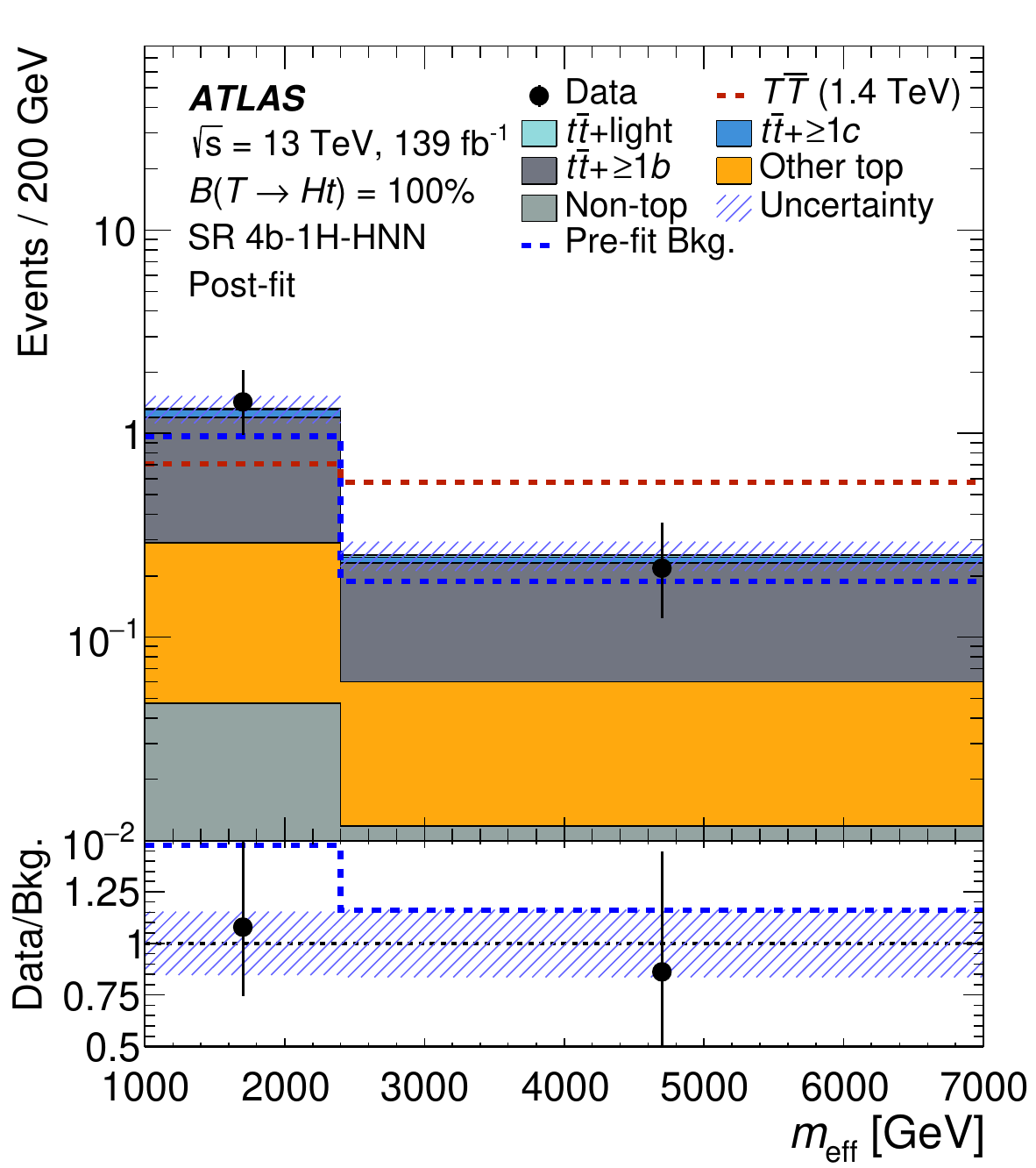}}
\caption{%
Comparison between observed yields (black dots) and expected SM yields (coloured areas) as a function of \mEff\ for the \HMVA\ signal regions (a) SR~$3b$-$0H$-\HMVA, (b) SR~$3b$-$1H$-\HMVA, (c) SR~$4b$-$0H$-\HMVA, and (d) SR~$4b$-$1H$-\HMVA.
The expected yields are shown after a fit under the background-only hypothesis in the control and signal regions.
The bottom panels display the ratio of data and total SM background.
The size of the combined statistical and systematic uncertainty in the background prediction is indicated by a blue hatched band, including all sources of systematic uncertainty described in Section~\ref{sec:uncertainties}.
In the upper panel, the total expected SM yield (expected yield from a \BTtoHtOne\ signal with $m_T=\SI{1.4}{TeV}$) before the fit is shown as a dashed blue (red) line.
In the lower panel, the ratio of the observed yield to the total expected SM yield before the fit is shown as a dashed blue line.
}
\label{fig:results_likelihood_mEff}
\end{figure}

Distributions as a function of \mEff\ for the analysis regions most sensitive to a hypothetical signal are shown in Figure~\ref{fig:results_likelihood_mEff}, with all remaining analysis regions delegated to Appendix~\ref{sec:appB}.
Good post-fit agreement between observed data and SM background prediction is obtained in all control and signal regions after the fit, including those in which the largest fraction of signal events is expected.

The impact of the different sources of uncertainty on the signal strength parameter~$\mu$ is summarised in Table~\ref{tab:results_likelihood_impact} for vector-like \TQuark{}s with $m_T=\SI{1.4}{TeV}$ in the \singlet, \doublet, and \BTtoHtOne\ scenarios.
The results are dominated by the statistical uncertainties on the data.
Among the systematic uncertainties, the largest contribution arises from the modelling of SM backgrounds, in particular \ttbar production.
The background reweighting uncertainty, although individually small, has a sizeable impact because its effect mirrors that of the signal: it is suppressed at low \mEff\ and enhanced at high \mEff, allowing it to either obscure or amplify signal contributions.
Considered alongside other modelling uncertainties, however, the background reweighting impacts are of comparable magnitude, indicating that the fit is robust against variations of the pre-fit background model.
Among experimental uncertainties, those related to jets have the largest impact.
While the absolute impact of the uncertainties varies between signal scenarios, their relative hierarchy remains similar.

\subsection{Exclusion limits on pair-production of vector-like quarks}

\begin{figure}[htbp!]
\centering
\subfloat[]{\includegraphics[width=0.49\textwidth]{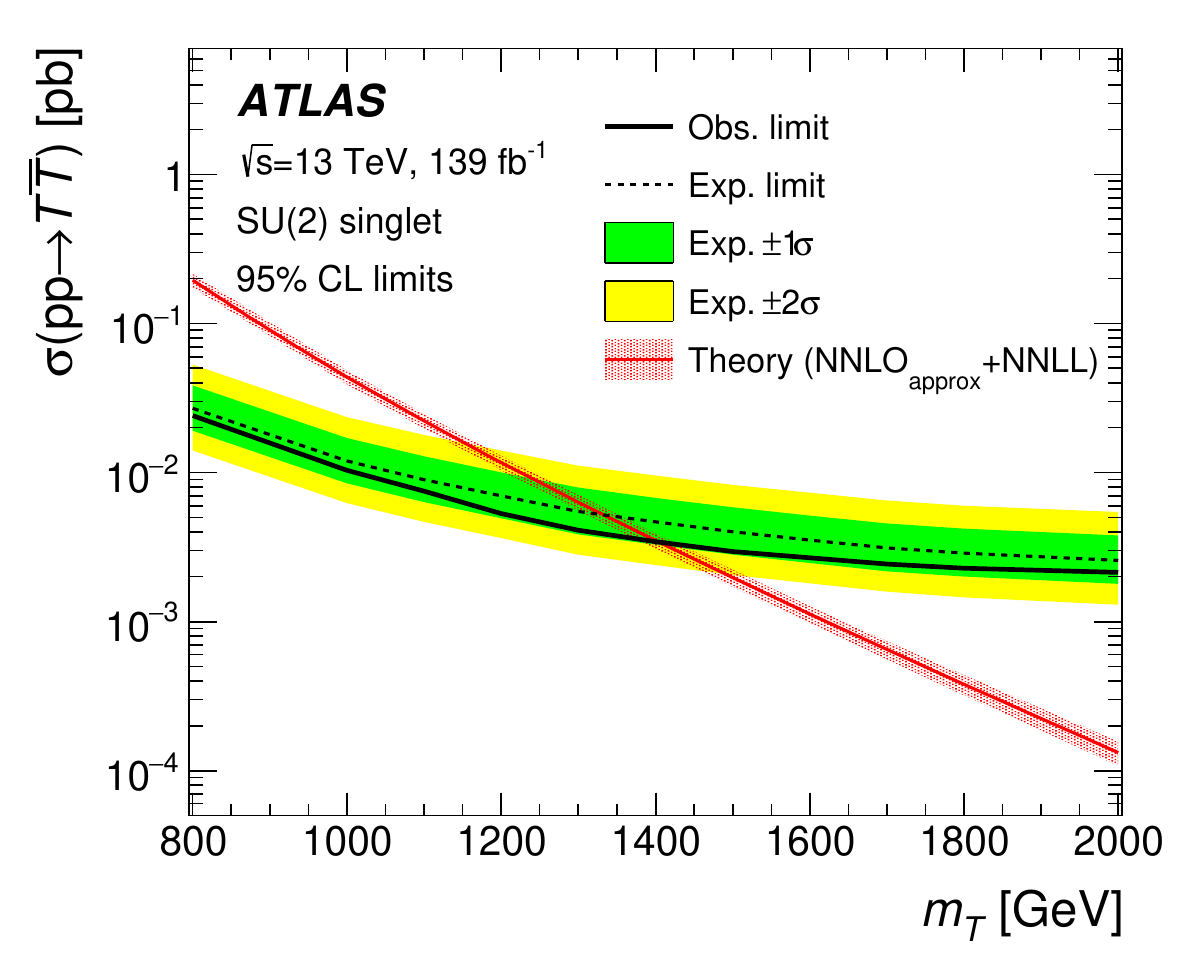}}
\subfloat[]{\includegraphics[width=0.49\textwidth]{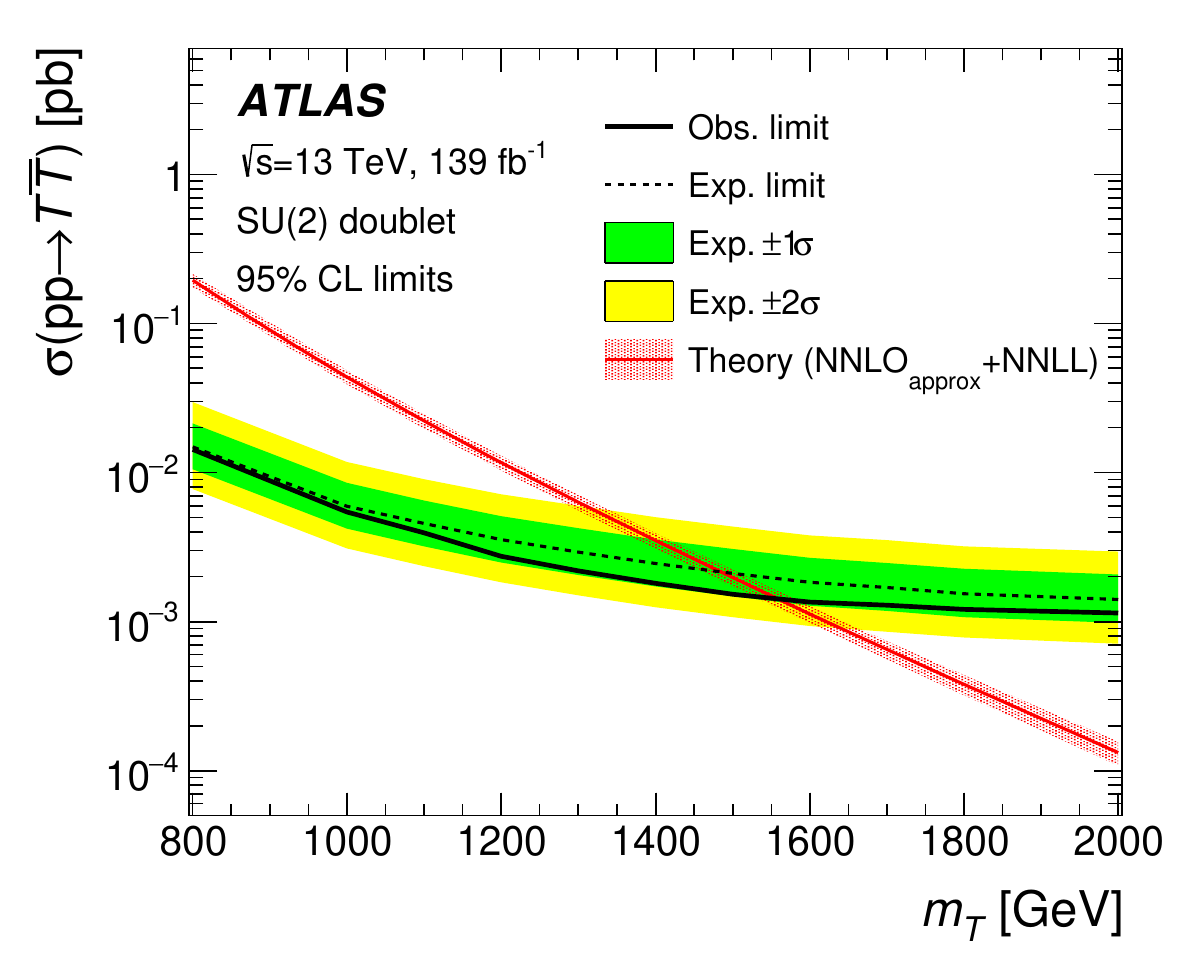}}\\
\subfloat[]{\includegraphics[width=0.49\textwidth]{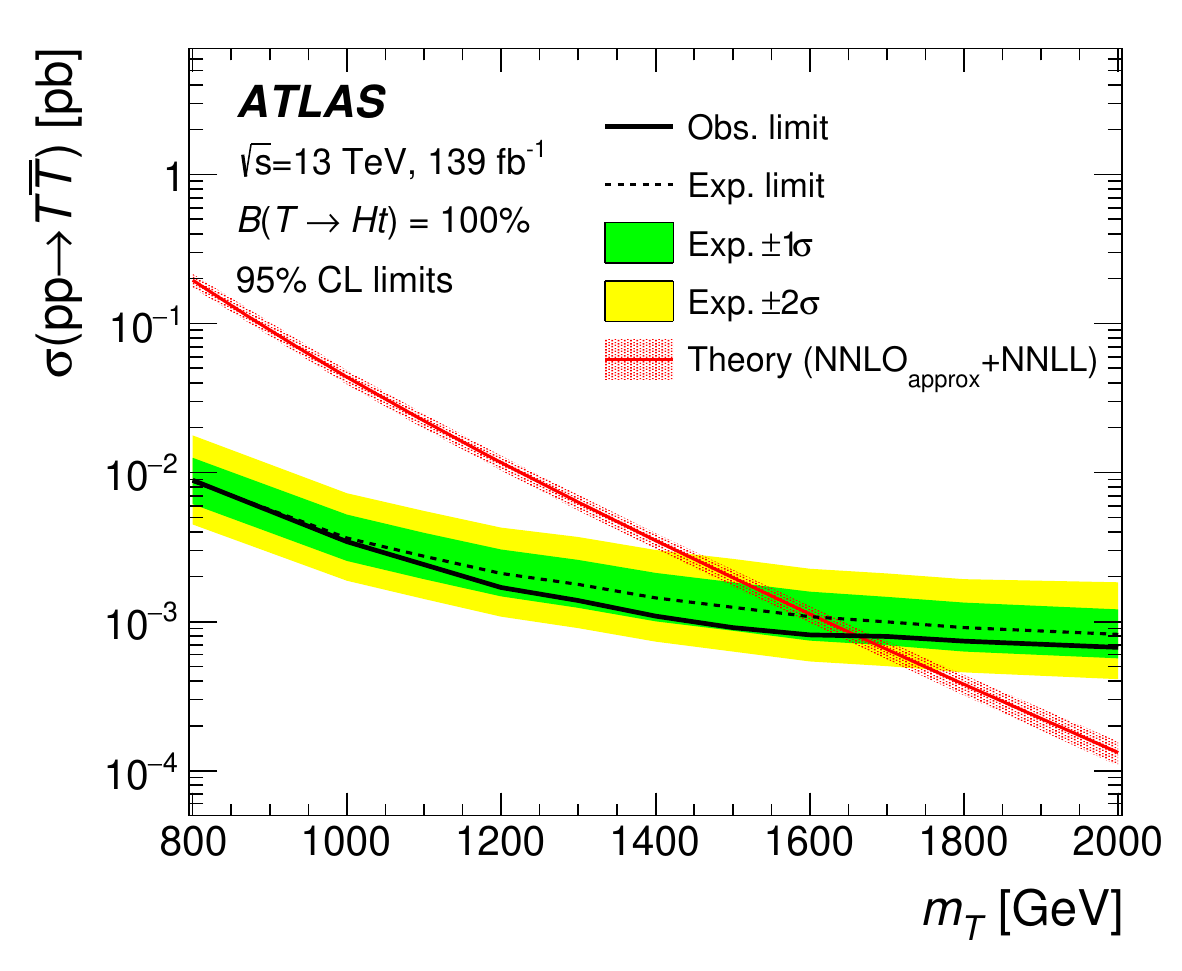}}
\caption{%
Observed and expected 95\% CL upper limits on the \TT\ pair-production cross-section as a function of the \TQuark\ mass for the (a) \singlet, (b) \doublet, and (c) \BTtoHtOne\ benchmark scenarios.
The inner green (outer yellow) band corresponds to the $\pm1\sigma$ ($\pm2\sigma$) uncertainty around the expected limits.
The theoretical cross-section prediction, and its corresponding uncertainty are also shown with a red line and light red band, respectively.
}
\label{fig:results_limits}
\end{figure}

Upper limits on the \TT\ production cross-section as a function of the \TQuark{} mass are derived.
These are shown in Figure~\ref{fig:results_limits} for the three different benchmark scenarios.
The observed (expected) $95\%$ \CL\ lower limits on the mass of pair-produced vector-like \TQuark{}s are \limitSingletUnit\ (\limitSingletExpUnit) in the \singlet, \limitDoubletUnit\ (\limitDoubletExpUnit) in the \doublet, and \limitHtHtUnit\ (\limitHtHtExpUnit) in the \BTtoHtOne\ scenario.
These are the most stringent limits in the \doublet\ and \BTtoHtOne\ scenarios to date, extending the exclusion by \improveDoubletCMS\ and \improveHtHtCMS, respectively, compared with the current strongest limits set by a CMS search~\cite{CMS-B2G-20-011}.
They are also the most stringent limits set by the ATLAS Collaboration for all three scenarios, extending the exclusion by \improveSingletATLAS\ in the \singlet\ scenario~\cite{EXOT-2019-06}, \improveDoubletATLAS\ in the \doublet\ scenario~\cite{EXOT-2018-58}, and \improveHtHtATLAS\ in the \BTtoHtOne\ scenario~\cite{EXOT-2016-13}.
An overview of the lower limits on the mass of pair-produced vector-like \TQuark{}s from this and related works is given in Table~\ref{tab:limits_overview}.

\begin{table}
\caption{%
Summary of observed (expected) 95\% \CL\ lower limits on the mass of pair-produced vector-like \TQuark{}s under the \singlet, \doublet, and \BTtoHtOne\ scenarios for this and other related results.
}
\label{tab:limits_overview}
\centering
\resizebox{\textwidth}{!}{%
\begin{tabular}{cccccc}
\toprule
\multicolumn{3}{c}{\textbf{Analysis}} & \multicolumn{3}{c}{\textbf{Exclusion limits [TeV]}}\\
\cmidrule(lr){1-3}\cmidrule(lr){4-6}
Reference & Experiment & Integrated luminosity [fb$^{-1}$] & \singlet & \doublet & \BTtoHtOne\\
\midrule
This result & ATLAS & \usedIntLumi & \limitSinglet\ (\limitSingletExp) & \limitDoublet\ (\limitDoubletExp) & \limitHtHt\ (\limitHtHtExp)\\
\midrule
\cite{EXOT-2016-13} & ATLAS & 36.1 & 1.19 (1.11) & 1.31 (1.26) & 1.43 (1.34)\\
\cite{EXOT-2017-17} & ATLAS & 36.1 & 1.31 (1.23) & 1.37 (1.32) & 1.43 (1.34)\\
\cite{EXOT-2018-58} & ATLAS & 139  & 1.27 (1.28) & 1.46 (1.44) & --\\
\cite{EXOT-2019-06} & ATLAS & 140  & 1.36 (1.36) & -- & -- \\
\cite{CMS-B2G-20-011} & CMS & 138  & 1.48 (1.41) & 1.49 (1.50) & 1.50 (1.55)\\
\bottomrule
\end{tabular}
}
\end{table}


%
%
%

\section{Conclusion}
\label{sec:conclusion}

A search for the pair production of up-type vector-like \TQuark{}s, where at least one \TQuark{} decays via the $T\to Ht$ channel with subsequent $H\to\bbbar$ decay, is presented.
The search uses \usedIntLumiUnit\ of $pp$ collisions at $\sqrt{s}=\SI{13}{TeV}$, collected by the ATLAS detector at the LHC during the 2015--2018 Run 2 data-taking period.
Events are analysed in the lepton-plus-jets final state, characterised by the presence of a single electron or muon, multiple jets and $b$-tagged jets in the event.
The search makes use of the presence of boosted, hadronically decaying Higgs and vector bosons, and hadronically or leptonically decaying top quarks in signal events.
No significant excess above the SM expectation is observed, and exclusion limits at $95\%$ \CL\ are set on the production cross-section of pair-produced \TQuark{}s.
In the \singlet\ (doublet) scenario, \TQuark{}s with masses below \limitSingletUnit\ (\limitDoubletUnit) are excluded.
If \BTtoHtOne, \TQuark{}s with masses below \limitHtHtUnit\ are excluded.
These are the most stringent limits in the \doublet\ and \BTtoHtOne\ scenarios to date.
As the search remains limited by statistical uncertainties, the substantially increased data from Run~3 and the High-Luminosity LHC offer significant potential for extending the sensitivity beyond the mass ranges probed here.
To facilitate reinterpretation and future studies, event yields, systematic uncertainties, exclusion limits, and the statistical workspace are provided on \textsc{HEPData}~\cite{Maguire:2017ypu}, along with the \NN~weights in both their original form and a portable format produced using \textsc{PetrifyML}~\cite{Buckley:2025pqk}.
A \textsc{Rivet}~\cite{Bierlich:2019rhm,Bierlich:2024vqo} analysis routine is also provided to enable direct comparison of the result with theoretical predictions.


%
%
%
%
%
\section*{Acknowledgements}
%

%

%
%

%
%

We thank CERN for the very successful operation of the LHC and its injectors, as well as the support staff at
CERN and at our institutions worldwide without whom ATLAS could not be operated efficiently.

The crucial computing support from all WLCG partners is acknowledged gratefully, in particular from CERN, the ATLAS Tier-1 facilities at TRIUMF/SFU (Canada), NDGF (Denmark, Norway, Sweden), CC-IN2P3 (France), KIT/GridKA (Germany), INFN-CNAF (Italy), NL-T1 (Netherlands), PIC (Spain), RAL (UK) and BNL (USA), the Tier-2 facilities worldwide and large non-WLCG resource providers. Major contributors of computing resources are listed in Ref.~\cite{ATL-SOFT-PUB-2026-001}.

We gratefully acknowledge the support of ANPCyT, Argentina; YerPhI, Armenia; ARC, Australia; BMWFW and FWF, Austria; ANAS, Azerbaijan; CNPq and FAPESP, Brazil; NSERC, NRC and CFI, Canada; CERN; ANID, Chile; CAS, MOST and NSFC, China; Minciencias, Colombia; MEYS CR, Czech Republic; DNRF and DNSRC, Denmark; IN2P3-CNRS and CEA-DRF/IRFU, France; SRNSFG, Georgia; BMFTR, HGF and MPG, Germany; GSRI, Greece; RGC and Hong Kong SAR, China; ICHEP and Academy of Sciences and Humanities, Israel; INFN, Italy; MEXT and JSPS, Japan; CNRST, Morocco; NWO, Netherlands; RCN, Norway; MNiSW, Poland; FCT, Portugal; MNE/IFA, Romania; MSTDI, Serbia; MSSR, Slovakia; ARIS and MVZI, Slovenia; DSI/NRF, South Africa; MICIU/AEI, Spain; SRC and Wallenberg Foundation, Sweden; SERI, SNSF and Cantons of Bern and Geneva, Switzerland; NSTC, Taipei; TENMAK, T\"urkiye; STFC/UKRI, United Kingdom; DOE and NSF, United States of America.

Individual groups and members have received support from BCKDF, CANARIE, CRC and DRAC, Canada; CERN-CZ, FORTE and PRIMUS, Czech Republic; COST, ERC, ERDF, Horizon 2020 and Marie Sk{\l}odowska-Curie Actions, European Union; Investissements d'Avenir Labex, Investissements d'Avenir Idex and ANR, France; DFG and AvH Foundation, Germany; Herakleitos, Thales and Aristeia programmes co-financed by EU-ESF and the Greek NSRF, Greece; BSF-NSF and MINERVA, Israel; NCN and NAWA, Poland; La Caixa Banking Foundation, CERCA and AGAUR programs from Generalitat de Catalunya and PROMETEO and GenT Programmes Generalitat Valenciana, Spain; G\"{o}ran Gustafssons Stiftelse, Sweden; The Royal Society and Leverhulme Trust, United Kingdom; Eric and Wendy Schmidt Fund for Strategic Innovation, United States of America.

In addition, individual members wish to acknowledge support from Chile: Agencia Nacional de Investigaci\'on y Desarrollo (ANID FONDECYT reg. 1230987, FONDECYT 1230812, FONDECYT 1240864, Fondecyt 3240661, Fondecyt Regular 1240721); China: Chinese Ministry of Science and Technology (MOST-2023YFA1605700, MOST-2023YFA1609300), National Natural Science Foundation of China (NSFC 12275265, NSFC-W2543005); Czech Republic: Czech Science Foundation (GACR - 24-11373S), Ministry of Education Youth and Sports (ERC-CZ-LL2327, FORTE CZ.02.01.01/00/22\_008/0004632), PRIMUS Research Programme (PRIMUS/21/SCI/017); EU: H2020 European Research Council (ERC - 101002463); European Union: European Research Council (BARD No. 101116429, ERC - 101219398, ERC - 948254, ERC 101089007), European Regional Development Fund (HE COFUND GA No.101081355, ERDF), Marie Sklodowska-Curie Actions (GAP-101168829); France: Agence Nationale de la Recherche (ANR-21-CE31-0013, ANR-22-EDIR-0002, ANR-24-CE31-0504-01); Germany: Deutsche Forschungsgemeinschaft (DFG - 469666862); China: Research Grants Council (GRF); Italy: Ministero dell'Università e della Ricerca (NextGenEU 153D23001490006 M4C2.1.1, NextGenEU I53D23000820006 M4C2.1.1, NextGenEU I53D23001490006 M4C2.1.1, SOE2024\_0000023); Japan: Japan Society for the Promotion of Science (JSPS KAKENHI  JP25H0063, JSPS KAKENHI JP22H01227, JSPS KAKENHI JP22H04944, JSPS KAKENHI JP22KK0227, JSPS KAKENHI JP24K23939, JSPS KAKENHI JP24KK0251, JSPS KAKENHI JP25H00650, JSPS KAKENHI JP25H01291, JSPS KAKENHI JP25K01011, JSPS KAKENHI JP25K01023); Norway: Research Council of Norway (RCN-314472); Poland: Polish National Science Centre (NCN 2021/42/E/ST2/00350, NCN OPUS 2023/51/B/ST2/02507, NCN OPUS nr 2022/47/B/ST2/03059, NCN UMO-2019/34/E/ST2/00393, UMO-2022/47/O/ST2/00148, UMO-2023/49/B/ST2/04085, UMO-2023/51/B/ST2/00920, UMO-2024/53/N/ST2/00869); Spain: Agència de Gestió d'Ajuts Universitaris i de Recerca. (AGAUR - 2023 BP 00141), Ministry of Science and Innovation (RYC2019-028510-I, RYC2020-030254-I, RYC2021-031273-I, RYC2022-038164-I), Ministerio de Ciencia, Innovación y Universidades/Agencia Estatal de Investigaci\'on (EU NextGenerationEU (PRTR-C17.I1), PID2022-142604OB-C22); Sweden: Carl Trygger Foundation (Carl Trygger Foundation CTS 22:2312), Swedish Research Council (Swedish Research Council 2023-04654, VR 2021-03651, VR 2022-03845, VR 2022-04683, VR 2023-03403, VR 2024-05451, VR 2025-05940), Knut and Alice Wallenberg Foundation (KAW 2023.0366); Switzerland: Swiss National Science Foundation (SNSF - PCEFP2\_194658); United Kingdom: The Binks Trust, Royal Society (NIF-R1-231091); United States of America: U.S. Department of Energy (ECA DE-AC02-76SF00515), John Templeton Foundation (John Templeton Foundation 63206), Neubauer Family Foundation.

%
%


%
\clearpage
\appendix
\part*{Appendix}
\addcontentsline{toc}{part}{Appendix}

\section{Complete list of neural network input variables}
\label{sec:appNNInputVariables}
\begin{table*}[!htbp]
\centering
\caption{Input variables for the multilayer perceptron ranked according to their importance and separation power.}
\label{tab:mva_vars}
\def\arraystretch{1.2}
\resizebox{\textwidth}{!}{%
\begin{tabular}{rcl}
\textbf{Number} & \textbf{Variable} & \textbf{Description}\\
\toprule
1.  & \mEff & effective mass: scalar sum of the \pT\ of all selected \smallR\ jets and the lepton as well as \MET\\ [-0.2em]
2.  & \objPt[3]{\varRShort} & \pT of third-leading \varR\ jet\\
3.  & \objPt{\tlreco} & \pT of \tlreco\\
4.  & \nVarRJets & number of \varR\ jets\\
5.  & \objPt[1]{H} & \pT of leading $H$-tagged jet\\
6.  & \DeltaPhiAvg[\varRShort] & average pairwise \DeltaPhi\ separation between all \varR\ jets\\
7.  & \mVLQ{\Tm[2]} & mass of the \TQuark\ candidate built from mass-tagged jets with the second-largest \DeltaR\ \\[-0.2em]
& & separation between its constituents\\
8.  & \nConst[1]{H} & number of constituents of leading $H$-tagged jet\\
9.  & \nBConst[1]{H} & number of $b$-tagged constituents of leading $H$-tagged jet\\
10. & \DeltaRMin[\mathrm{mass}] & minimum pairwise \DeltaR\ separation between all mass-tagged jets\\
\midrule
11. & \DeltaRMin[\varRShort] & minimum pairwise \DeltaR\ separation between all \varR\ jets\\
12. & \mVLQ{\TvarR[2]} & mass of the \TQuark\ candidate built from \varR\ jets with the second-largest \DeltaR\ \\ [-0.2em]
& & separation between its constituents\\
13. & \mTbMin & minimum transverse mass between \met and any of the three leading $b$-tagged jets\\
14. & \met & missing transverse energy\\
15. & \nBJets & number of $b$-tagged jets\\
16. & \DeltaPhiMin[\mathrm{mass}] & minimum pairwise \DeltaPhi\ separation between all mass-tagged jets\\
17. & \DeltaRAvg[\varRShort] & average pairwise \DeltaR\ separation between all \varR\ jets\\
18. & \DeltaPhiMin[\varRShort] & minimum pairwise \DeltaPhi\ separation between all \varR\ jets\\
19. & \nConst[1]{V} & number of constituents of leading $V$-tagged jet\\
20. & \objPt[1]{V} & \pT of leading $V$-tagged jet\\
\midrule
21. & \metRes & residual \met: difference between \met and the neutrino \pT as determined in the\\ [-0.2em]
& & \Wlreco\ reconstruction\\
22. & \DeltaEtaMin[\mathrm{mass}] & minimum pairwise \DeltaEta\ separation between all mass-tagged jets\\
23. & \objPt[1]{t} & \pT of leading $t$-tagged jet\\
24. & \mVLQ{\Tm[1]} & mass of the \TQuark\ candidate built from mass-tagged jets with the largest \DeltaR\ separation\\ [-0.2em]
& & between its constituents\\
25. & \DeltaEtaMin[\varRShort] & minimum pairwise \DeltaEta\ separation between all \varR\ jets\\
26. & \DeltaEtaLead[\varRShort] & \DeltaEta\ separation between two leading \varR\ jets\\
27. & \mTW & transverse mass of lepton--\pTmiss system\\
28. & \DeltaRLead[\varRShort] & \DeltaR\ separation between two leading \varR\ jets\\
29. & \objPt[2]{H} & \pT of subleading $H$-tagged jet\\
30. & \DeltaRLead[\mathrm{mass}] & \DeltaR\ separation between two leading mass-tagged jets\\
\bottomrule
\end{tabular}}
\end{table*}
\FloatBarrier


\clearpage

\section{Complete set of data and predicted yields in analysis regions}
\label{sec:appB}

\begin{figure}
\centering
\subfloat[]{\includegraphics[width=0.4\textwidth]{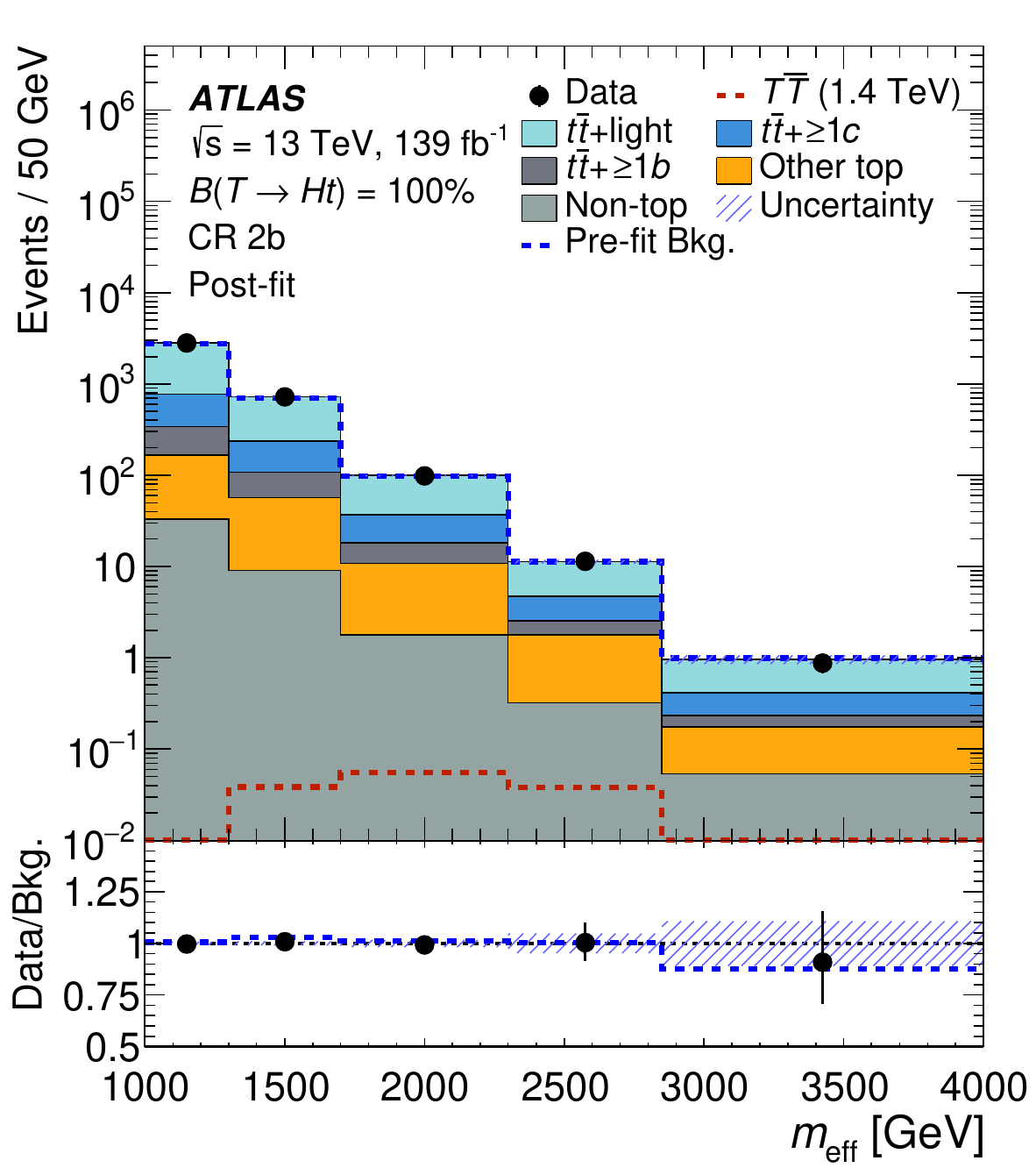}}\\
\subfloat[]{\includegraphics[width=0.4\textwidth]{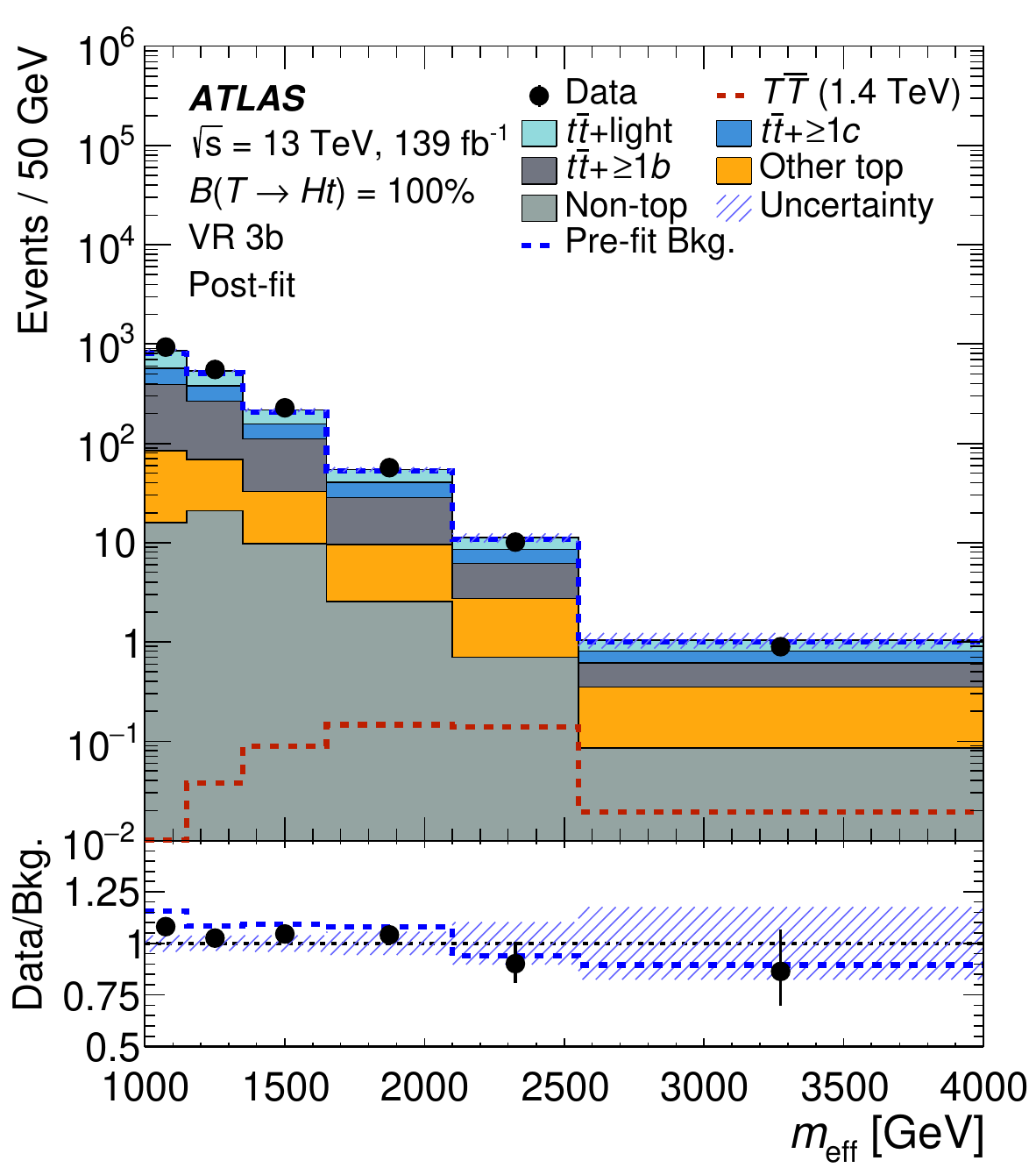}}
\subfloat[]{\includegraphics[width=0.4\textwidth]{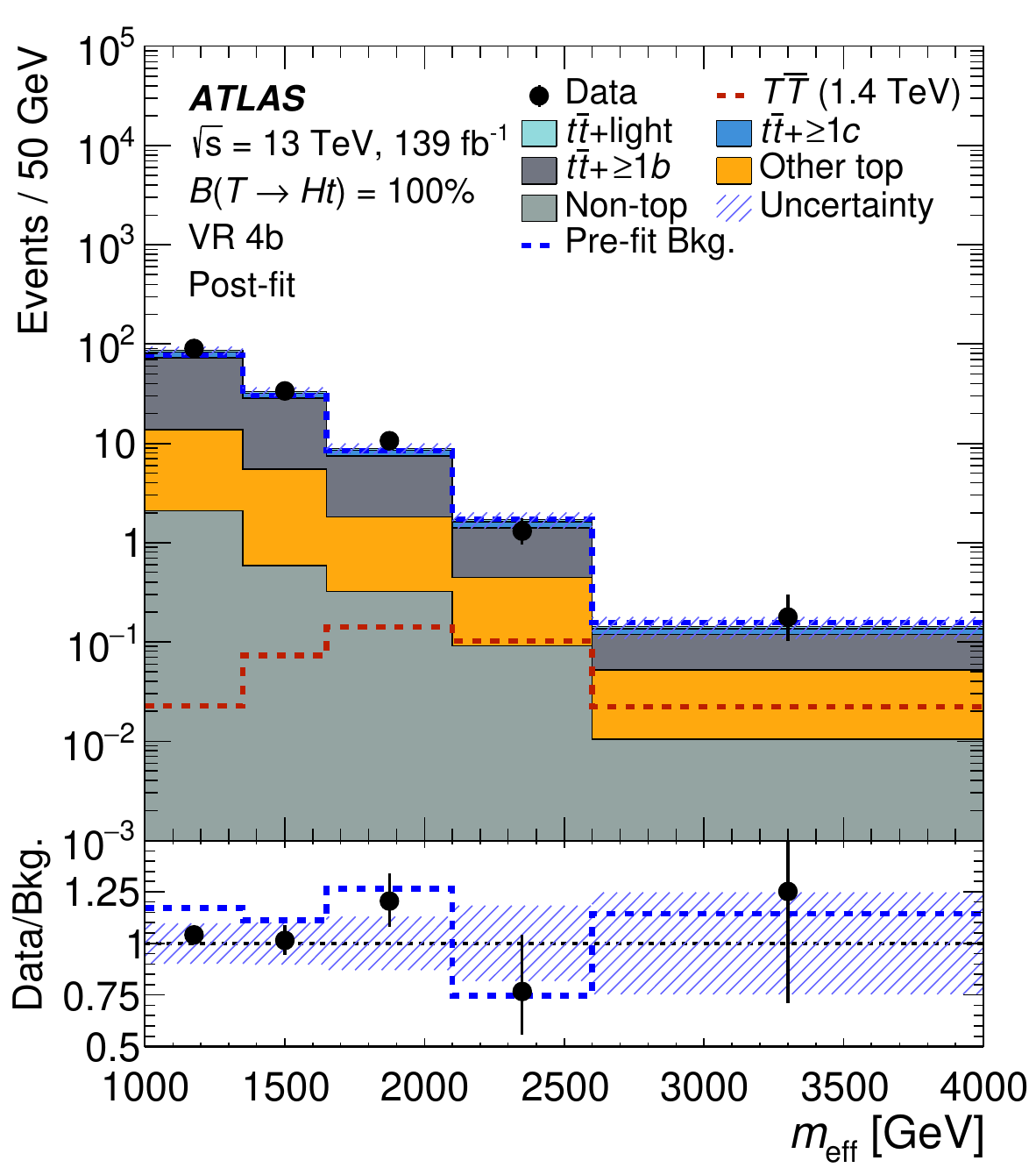}}
\caption{\mEffFigureCompactCaption{(a) the control region CR~$2b$ and the validation regions (b) VR~$3b$ and (c) VR~$4b$%
}}
\label{fig:results_likelihood_mEff_ttCRs}
\end{figure}

\begin{figure}
\centering
\subfloat[]{\includegraphics[width=0.4\textwidth]{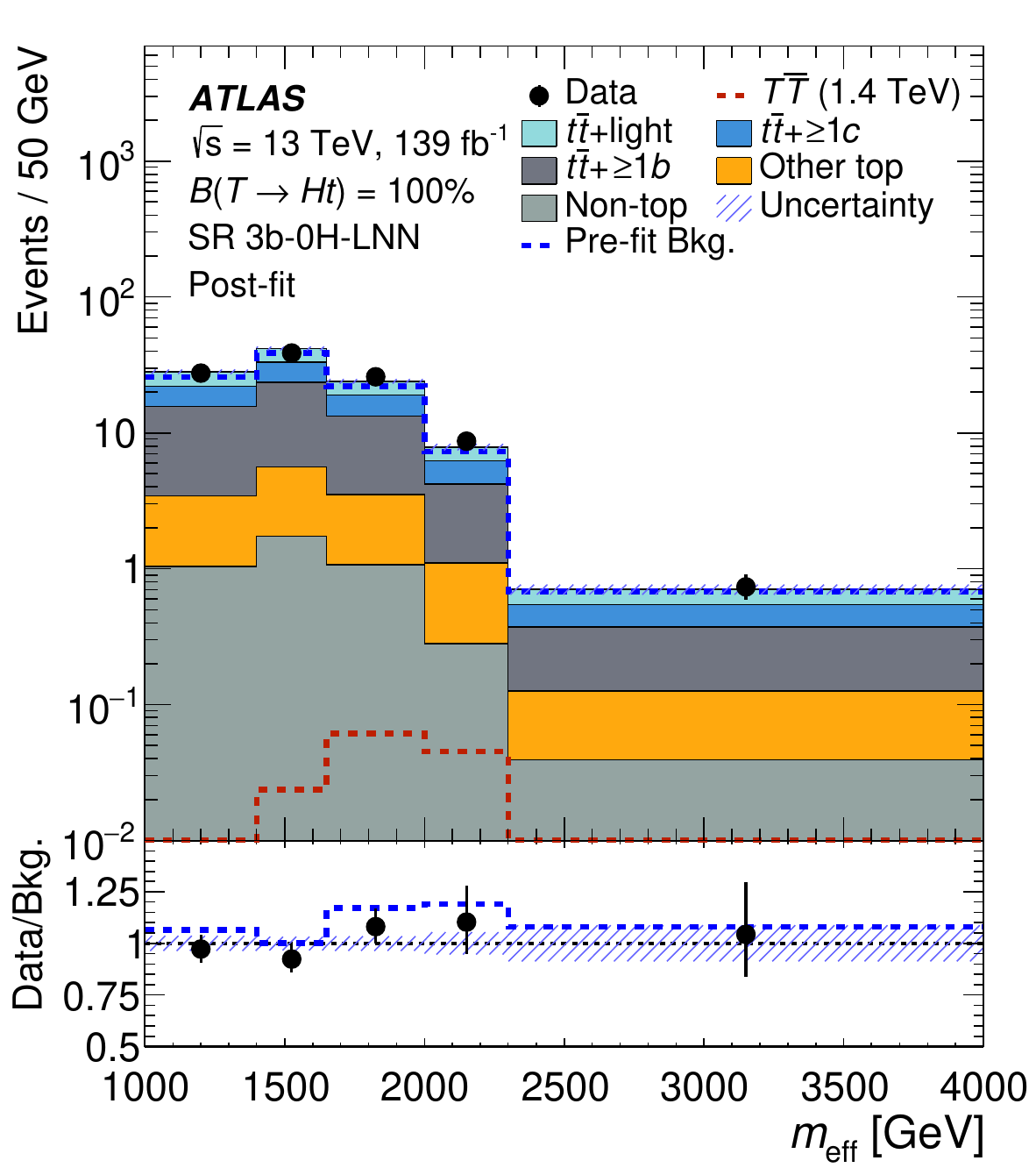}}
\subfloat[]{\includegraphics[width=0.4\textwidth]{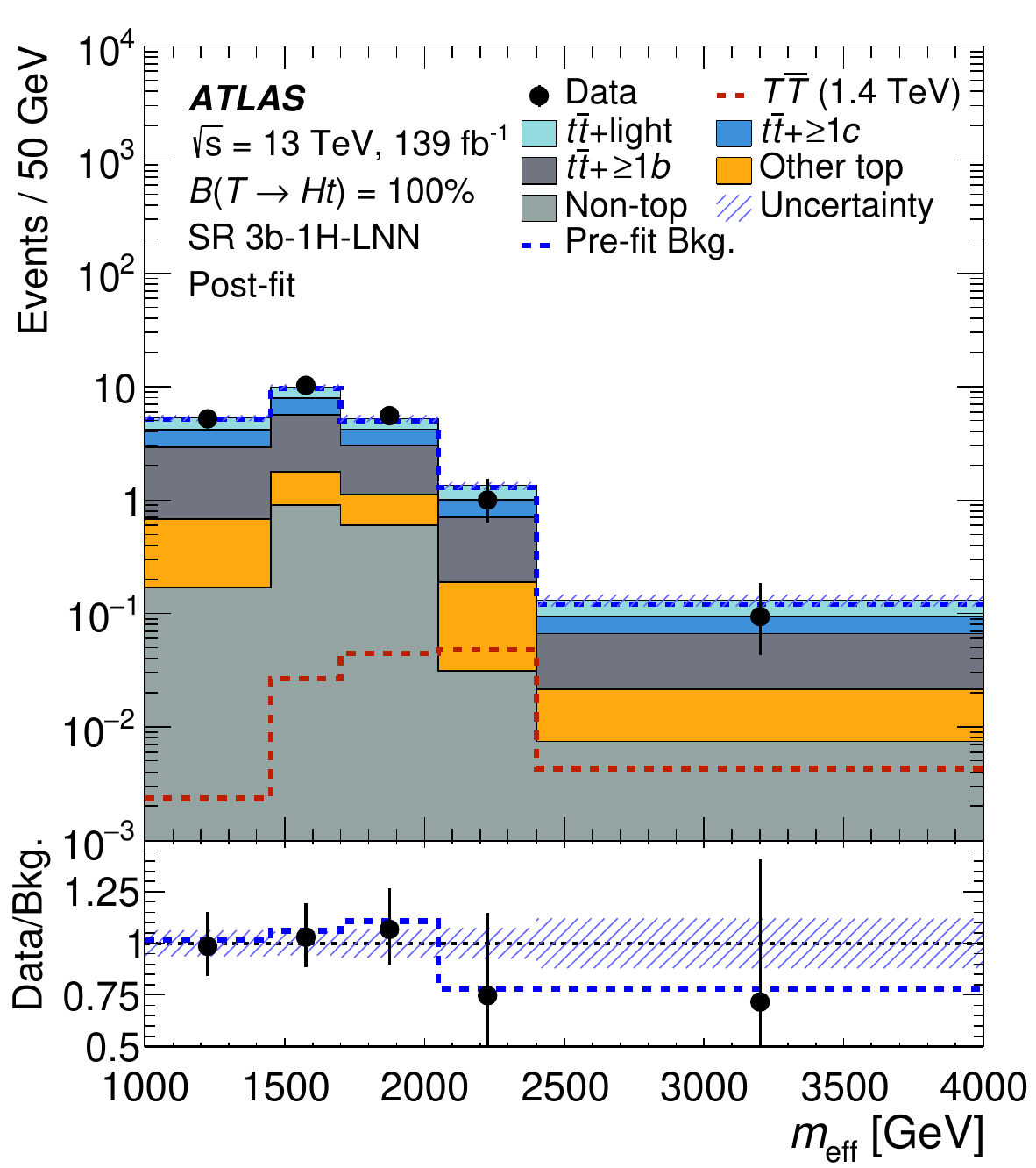}}\\
\subfloat[]{\includegraphics[width=0.4\textwidth]{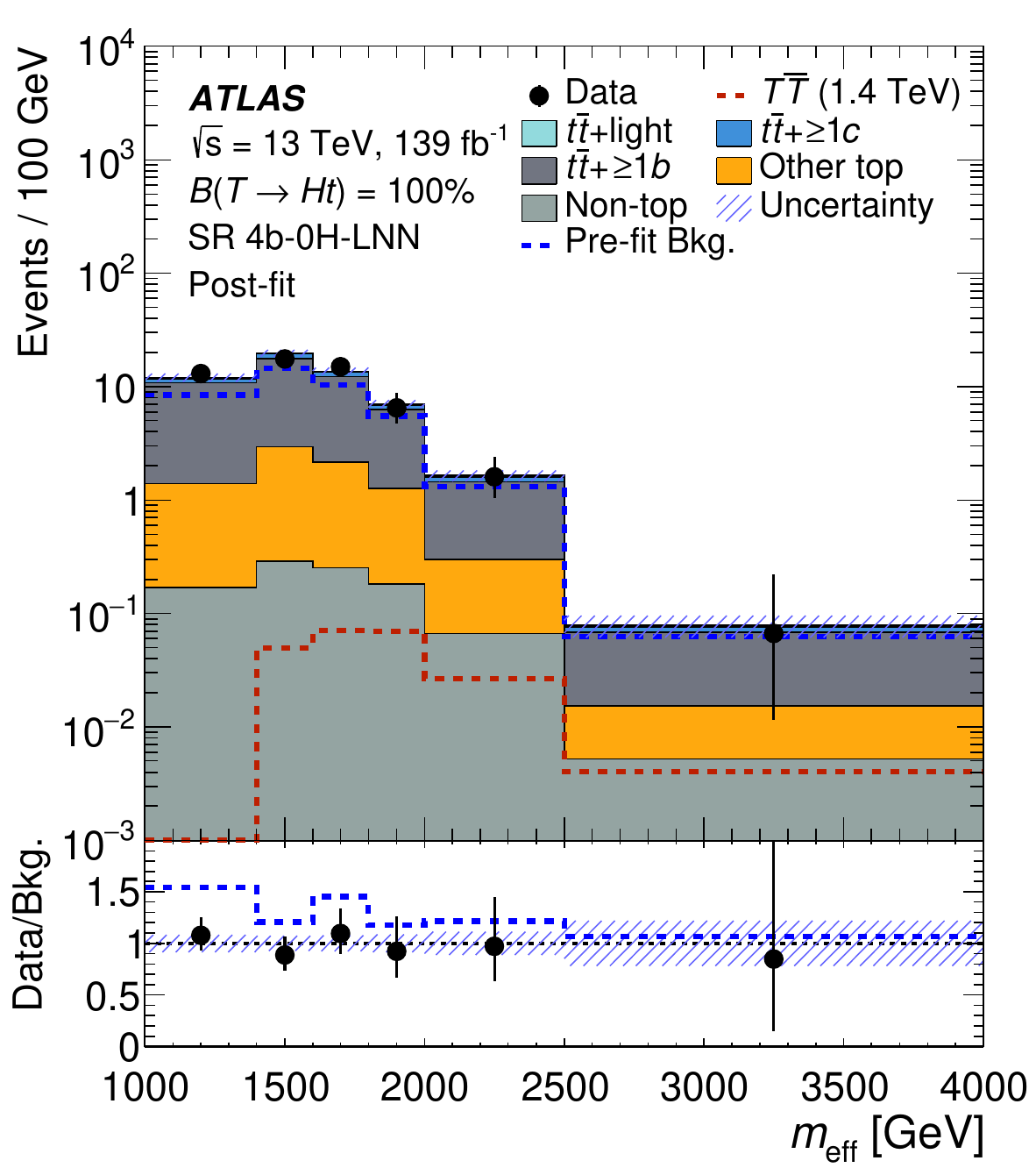}}
\subfloat[]{\includegraphics[width=0.4\textwidth]{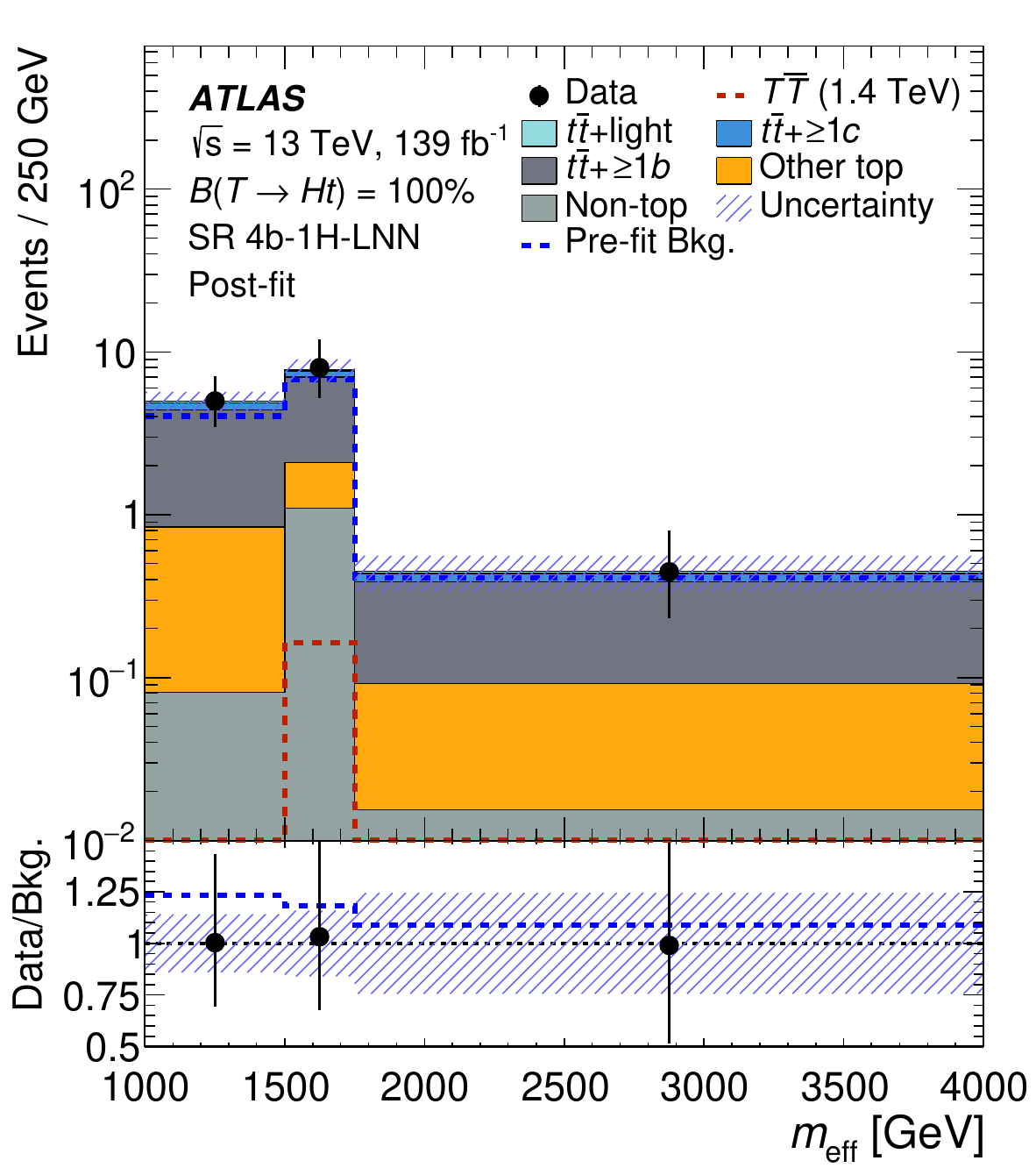}}
\caption{\mEffFigureCompactCaption{the \LMVA\ signal regions (a) SR~$3b$-$0H$-\LMVA, (b) SR~$3b$-$1H$-\LMVA, (c) SR~$4b$-$0H$-\LMVA, and (d) SR~$4b$-$1H$-\LMVA%
}}
\label{fig:results_likelihood_mEff_LMVA}
\end{figure}

\begin{figure}
\centering
\subfloat[]{\includegraphics[width=0.4\textwidth]{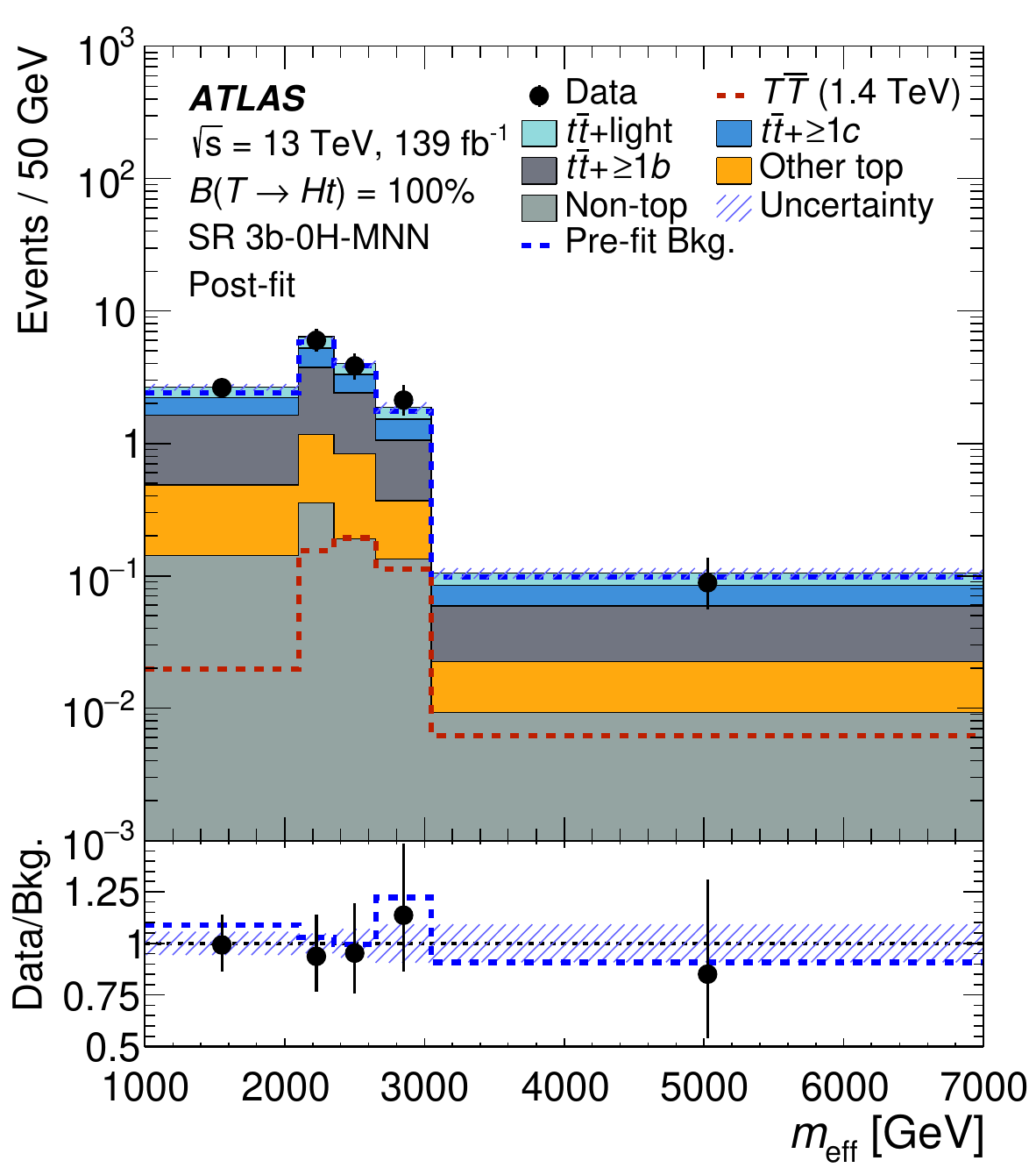}}
\subfloat[]{\includegraphics[width=0.4\textwidth]{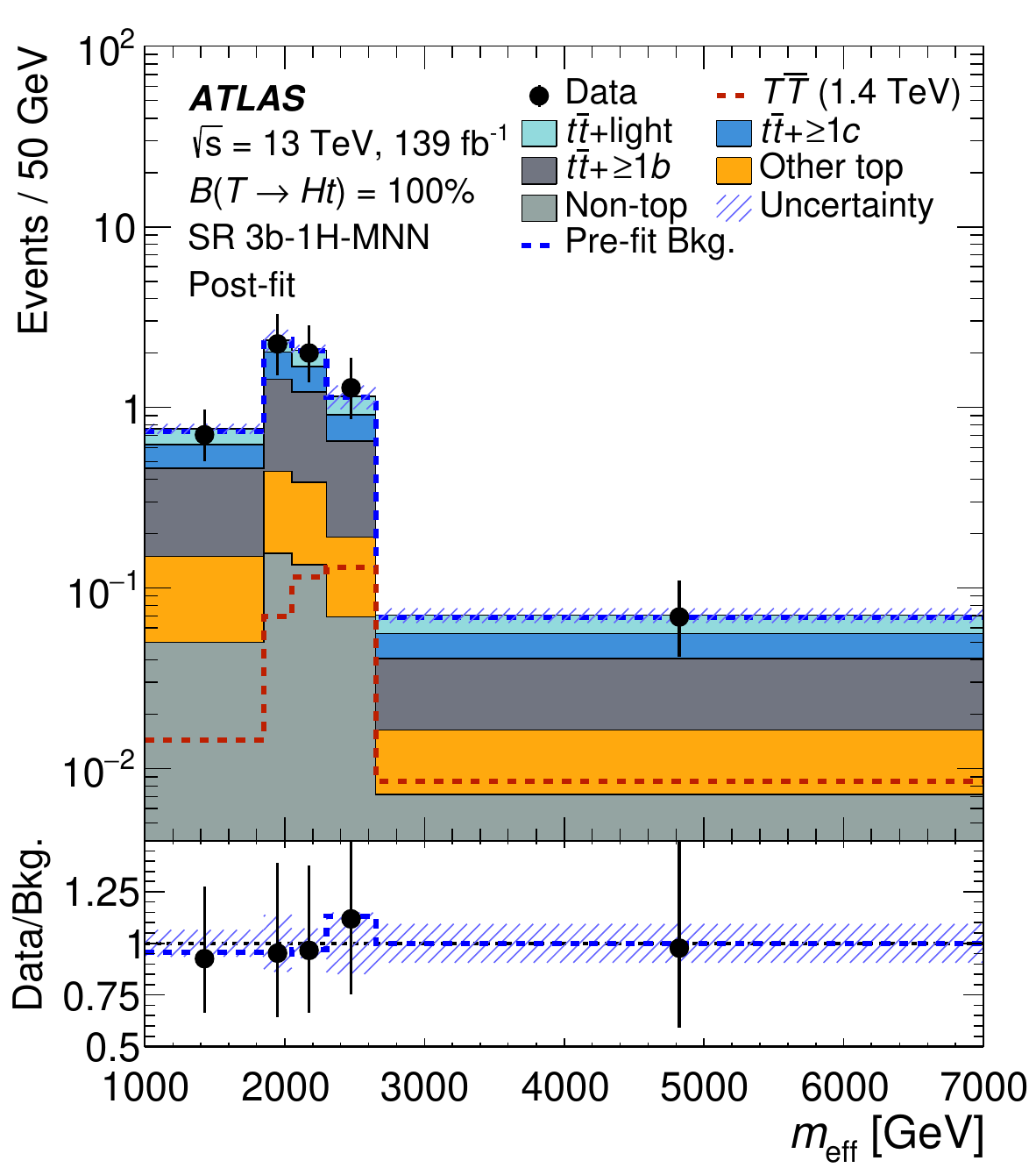}}\\
\subfloat[]{\includegraphics[width=0.4\textwidth]{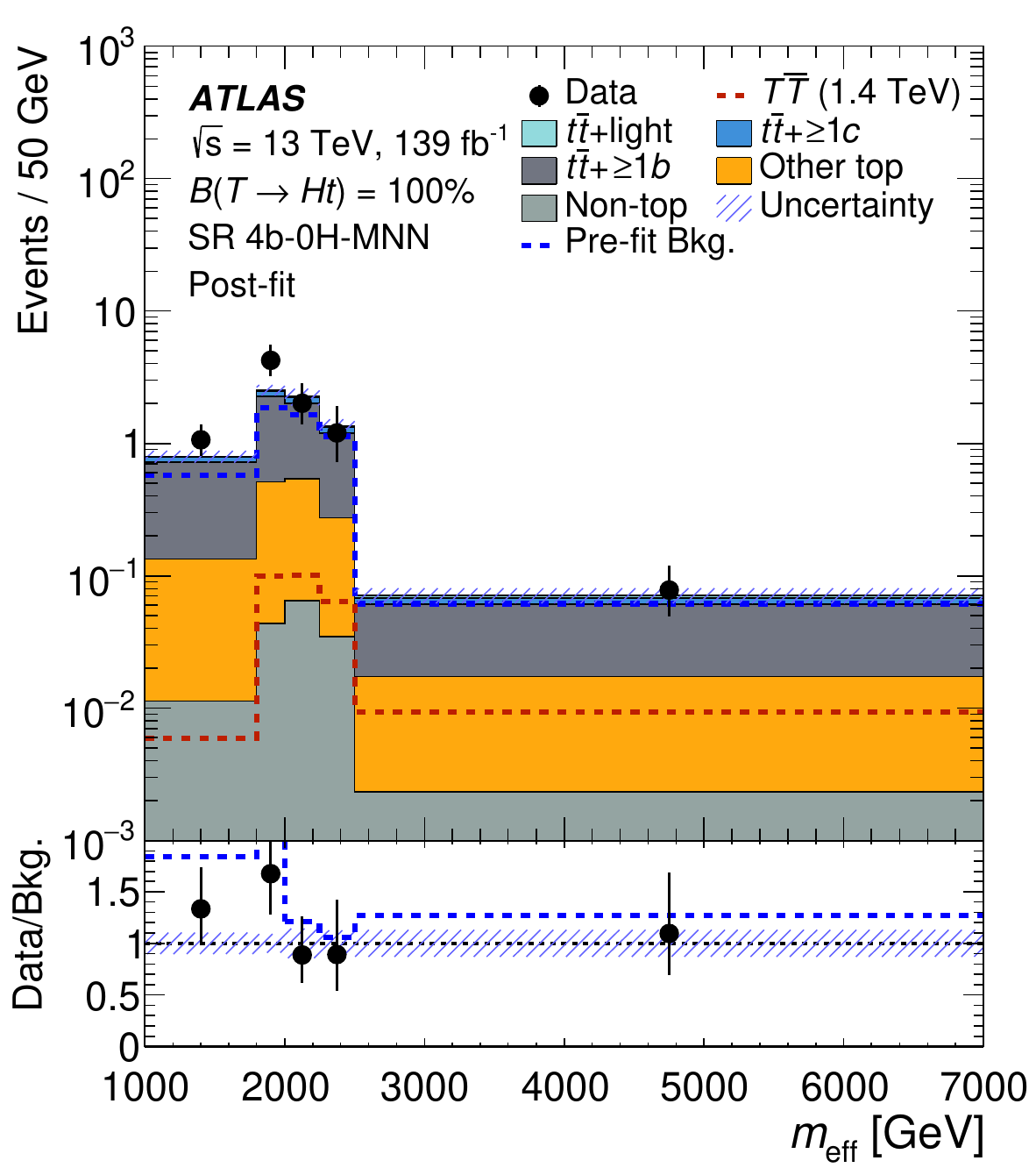}}
\subfloat[]{\includegraphics[width=0.4\textwidth]{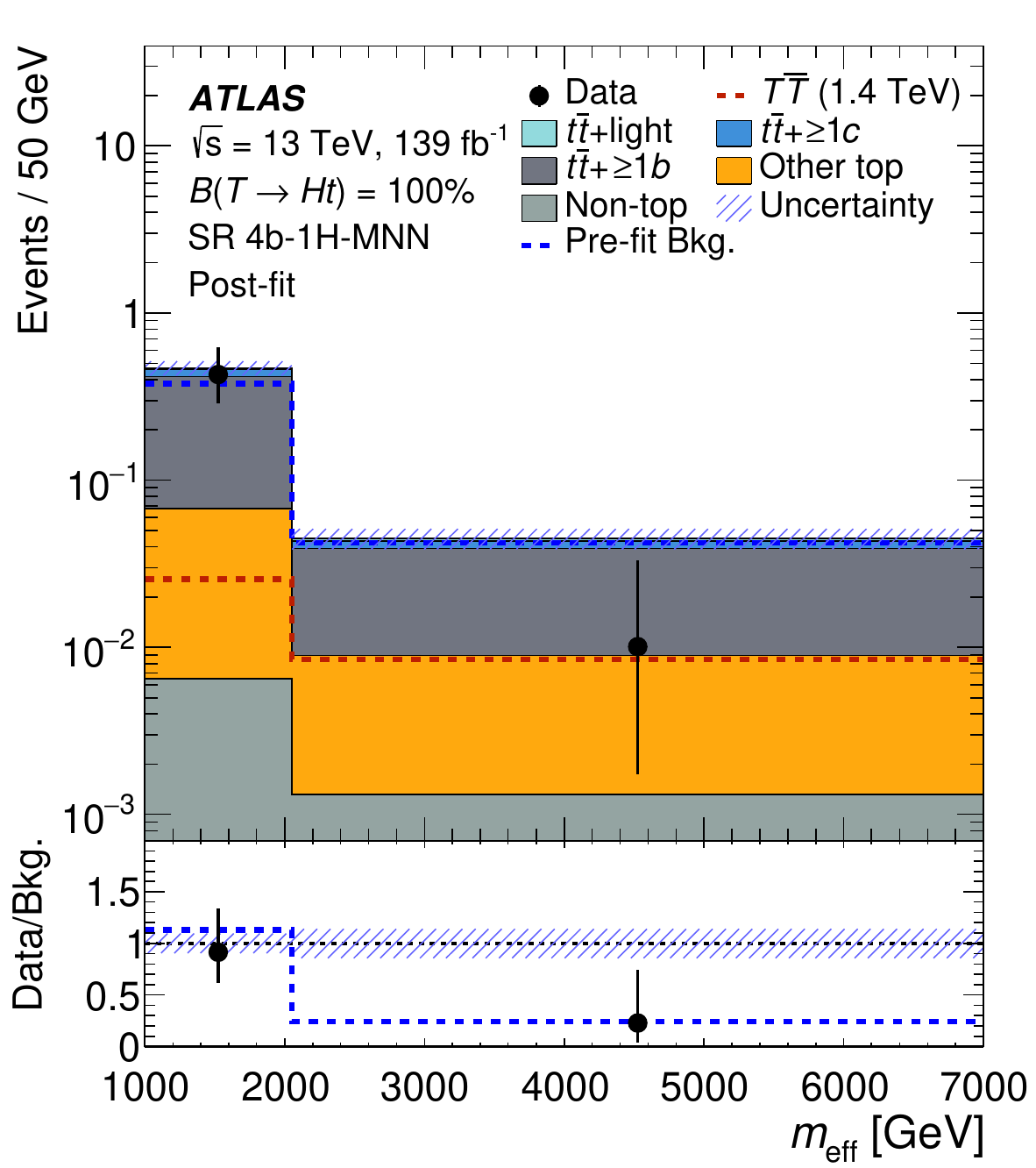}}
\caption{\mEffFigureCompactCaption{the \MMVA\ signal regions (a) SR~$3b$-$0H$-\MMVA, (b) SR~$3b$-$1H$-\MMVA, (c) SR~$4b$-$0H$-\MMVA, and (d) SR~$4b$-$1H$-\MMVA%
}}
\label{fig:results_likelihood_mEff_MMVA}
\end{figure}

A full set of tables with the expected and observed yields in all control, validation and signal regions, after a background-only fit to data in the control and signal regions, is provided in
Tables~\ref{tab:yields_CR_after} to \ref{tab:yields_SRHNN_after}.
To complete the results in Section~\ref{sec:results_likelihoodFit}, Figures~\ref{fig:results_likelihood_mEff_ttCRs} to \ref{fig:results_likelihood_mEff_MMVA} show the pre- and post-fit distributions of data and predicted yields as a function of \mEff\ in the remaining regions used in the search.

\FloatBarrier

\CRYieldTable{after}{%
\yieldRow{1.81 \pm 0.17  & 3.47 \pm 0.27 & 1.51 \pm 0.28}
\yieldRow{2.09 \pm 0.24  & 3.4 \pm 0.4   & 2.3 \pm 0.4}
\yieldRow{27   \pm 4     & 86 \pm 13     & 91 \pm 9}
\yieldRow{1.53 \pm 0.22  & 3.8 \pm 0.4   & 3.5 \pm 0.8}
\yieldRow{0.08 \pm 0.03  & 0.26 \pm 0.06 & 0.23 \pm 0.05}
\yieldRow{16900 \pm 1300 & 2010 \pm 190  &  34 \pm 4}
\yieldRow{3900 \pm 1500  & 1400 \pm 600  & 100 \pm 40}
\yieldRow{1570 \pm 300   & 2400 \pm 400  & 620 \pm 110}
\yieldRow{920 \pm 110    & 340  \pm 40   & 32  \pm 11}
\yieldRow{390 \pm 50     & 286 \pm 28    & 97  \pm 9}
\yieldRow{128 \pm 10     & 120 \pm 40    & 9.9 \pm 3.2}
\yieldRow{25.7 \pm 2.3   & 17 \pm 5      & 1.7 \pm 0.6}
\yieldRow{27.4 \pm 1.8   & 19 \pm 11     & 2.2 \pm 1.3}
\yieldRow{115  \pm 13    & 65 \pm 7      & 8.4 \pm 0.9}
\yieldRow{23950 \pm 160  & 6650 \pm 240  & 900 \pm 90}
\yieldRow{23958          & 6982          & 945}
}

\SRYieldTable{after}{\LMVA}{SRLNN}{%
\yieldRow{0.62 \pm 0.07 & 0.17 \pm 0.03 & 0.30 \pm 0.05 & 0.04 \pm 0.02}
\yieldRow{0.92 \pm 0.13 & 0.37 \pm 0.10 & 0.49 \pm 0.09 & 0.11 \pm 0.05}
\yieldRow{43 \pm 5      & 32.3 \pm 2.8  & 23 \pm 4      & 11.7 \pm 2.5}
\yieldRow{1.1 \pm 0.3   & 0.57 \pm 0.15 & 1 \pm 0.2     & 0.2  \pm 0.1}
\yieldRow{0.09 \pm 0.04 & 0.05 \pm 0.01 & 0.09 \pm 0.02 & 0.01 \pm 0.01}
\yieldRow{141 \pm 18    & 2.8  \pm 0.7  & 30.6 \pm 3.0  & 0.5  \pm 0.1}
\yieldRow{160 \pm 60    & 12   \pm 5    & 34 \pm 13     & 2.0  \pm 0.8}
\yieldRow{280 \pm 50    & 104  \pm 11   & 58 \pm 11     & 14.7 \pm 2.1}
\yieldRow{24  \pm 8     & 3.9  \pm 0.7  & 5.5  \pm 1.2  & 0.77 \pm 0.28}
\yieldRow{40 \pm 4      & 13.6 \pm 1.7  & 8.6  \pm 1.1  & 2.43 \pm 0.33}
\yieldRow{15 \pm 5      & 1.7 \pm 0.6   & 3.6 \pm 1.2   & 0.40 \pm 0.18}
\yieldRow{2.2 \pm 0.7   & 0.27 \pm 0.09 & 0.67 \pm 0.23 & 0.05 \pm 0.02}
\yieldRow{2.4 \pm 1.4   & 0.37 \pm 0.22 & 0.6 \pm 0.4   & 0.04 \pm 0.04}
\yieldRow{8.2 \pm 0.9   & 0.22 \pm 0.03 & 5.8 \pm 0.7   & 0.91 \pm 0.15}
\yieldRow{673 \pm 21    & 138 \pm 10    & 147 \pm 8     & 22 \pm 2}
\yieldRow{670           & 139           & 147           & 22}
}

\SRYieldTable{after}{\MMVA}{SRMNN}{%
\yieldRow{2.28 \pm 0.25 & 0.75 \pm 0.11 & 0.97 \pm 0.10 & 0.29 \pm 0.04}
\yieldRow{3.40 \pm 0.33 & 1.49 \pm 0.31 & 1.63 \pm 0.22 & 0.66 \pm 0.11}
\yieldRow{41 \pm 7      & 44 \pm 7      & 36 \pm 8      & 28.9 \pm 3.4}
\yieldRow{3.8 \pm 1.0   & 2.2 \pm 0.8   & 2.7 \pm 0.5   & 1.4 \pm 0.3}
\yieldRow{0.33 \pm 0.13 & 0.18 \pm 0.05 & 0.26 \pm 0.04 & 0.09 \pm 0.04}
\yieldRow{24.1 \pm 3.0  & 1.01 \pm 0.23 & 8.6 \pm 0.8   & 0.34 \pm 0.08}
\yieldRow{31 \pm 12     & 4.2 \pm 1.7   & 11 \pm 4      & 1.4 \pm 0.5}
\yieldRow{56 \pm 10     & 32 \pm 4      & 19 \pm 4      & 10.3 \pm 1.3}
\yieldRow{6.8 \pm 3.5   & 2.5 \pm 3.2   & 2.3 \pm 0.6   & 0.43 \pm 0.22}
\yieldRow{11.5 \pm 1.5  & 6.3 \pm 1.0   & 3.4 \pm 0.5   & 1.61 \pm 0.27}
\yieldRow{4.6 \pm 1.6   & 0.68 \pm 0.23 & 2.2 \pm 0.7   & 0.21 \pm 0.14}
\yieldRow{0.58 \pm 0.19 & 0.12 \pm 0.04 & 0.27 \pm 0.09 & 0.04 \pm 0.02}
\yieldRow{1.2 \pm 0.7   & 0.14 \pm 0.09 & 0.54 \pm 0.32 & 0.02 \pm 0.02}
\yieldRow{1.47 \pm 0.16 & 0.13 \pm 0.01 & 0.24 \pm 0.03 & \multicolumn{1}{c}{--}}
\yieldRow{138 \pm 7     & 47 \pm 4      & 47 \pm 3      & 14 \pm 1}
\yieldRow{135           & 57            & 46            & 10}
}

\SRYieldTable{after}{\HMVA}{SRHNN}{%
\yieldRow{2.90 \pm 0.34 & 2.6 \pm 0.4   & 4.03 \pm 0.34 & 3.9 \pm 0.5}
\yieldRow{3.8 \pm 0.7   & 5.2 \pm 0.9   & 6.7 \pm 0.7   & 9.1 \pm 1.2}
\yieldRow{20 \pm 5      & 52 \pm 12     & 49 \pm 9      & 106 \pm 14}
\yieldRow{3.3 \pm 1.2   & 6.8 \pm 2.1   & 10.5 \pm 1.8  & 18.2 \pm 2.9}
\yieldRow{0.44 \pm 0.16 & 0.67 \pm 0.31 & 1.65 \pm 0.30 & 2.43 \pm 0.34}
\yieldRow{2.3 \pm 0.4   & 0.37 \pm 0.14 & 3.1 \pm 0.5   & 0.25 \pm 0.06}
\yieldRow{3.1 \pm 1.2   & 1.4 \pm 0.6   & 3.9 \pm 1.5   & 1.1 \pm 0.6}
\yieldRow{6.2 \pm 1.2   & 11.3 \pm 1.6  & 7.7 \pm 1.5   & 10.3 \pm 1.9}
\yieldRow{2.6 \pm 1.1   & 0.56 \pm 0.31 & 2.4 \pm 0.5   & 0.78 \pm 0.31}
\yieldRow{2.4 \pm 0.4   & 2.7 \pm 0.6   & 2.3 \pm 0.3   & 2.0 \pm 0.4}
\yieldRow{0.95 \pm 0.35 & 0.39 \pm 0.13 & 1.1 \pm 0.4   & 0.31 \pm 0.13}
\yieldRow{0.12 \pm 0.05 & 0.04 \pm 0.02 & 0.11 \pm 0.05 & 0.03 \pm 0.02}
\yieldRow{0.14 \pm 0.18 & 0.08 \pm 0.05 & 0.21 \pm 0.13 & 0.09 \pm 0.06}
\yieldRow{0.15 \pm 0.02 & 0.36 \pm 0.05 & 0.19 \pm 0.02 & 0.16 \pm 0.02}
\yieldRow{18 \pm 2      & 17 \pm 2      & 21 \pm 1      & 15 \pm 2}
\yieldRow{16            & 15 			& 21 			& 15}
}


\FloatBarrier

\FloatBarrier


%
%
%
\printbibliography
\clearpage
 
\begin{flushleft}
\hypersetup{urlcolor=black}
{\Large The ATLAS Collaboration}

\bigskip

\AtlasOrcid[0000-0002-6665-4934]{G.~Aad}$^\textrm{\scriptsize 102}$,
\AtlasOrcid[0000-0001-7616-1554]{E.~Aakvaag}$^\textrm{\scriptsize 17}$,
\AtlasOrcid[0000-0002-5888-2734]{B.~Abbott}$^\textrm{\scriptsize 121}$,
\AtlasOrcid[0000-0002-0287-5869]{S.~Abdelhameed}$^\textrm{\scriptsize 83b}$,
\AtlasOrcid[0000-0002-1002-1652]{K.~Abeling}$^\textrm{\scriptsize 54}$,
\AtlasOrcid[0000-0001-5763-2760]{N.J.~Abicht}$^\textrm{\scriptsize 48}$,
\AtlasOrcid[0000-0002-8496-9294]{S.H.~Abidi}$^\textrm{\scriptsize 30}$,
\AtlasOrcid[0009-0003-6578-220X]{M.~Aboelela}$^\textrm{\scriptsize 44}$,
\AtlasOrcid[0000-0002-9987-2292]{A.~Aboulhorma}$^\textrm{\scriptsize 36e}$,
\AtlasOrcid[0000-0001-5329-6640]{H.~Abramowicz}$^\textrm{\scriptsize 154}$
% ...
\bigskip
\\

% ...
$^{17}$Department for Physics and Technology, University of Bergen, Bergen; Norway.\\
% ...
$^{30}$Physics Department, Brookhaven National Laboratory, Upton NY; United States of America.\\
% ...
$^{36}$$^{(a)}$Facult\'e des Sciences Ain Chock, Universit\'e Hassan II de Casablanca;$^{(b)}$Facult\'{e} des Sciences, Universit\'{e} Ibn-Tofail, K\'{e}nitra;$^{(c)}$Facult\'e des Sciences Semlalia, Universit\'e Cadi Ayyad, LPHEA-Marrakech;$^{(d)}$LPMR, Facult\'e des Sciences, Universit\'e Mohamed Premier, Oujda;$^{(e)}$Facult\'e des sciences, Universit\'e Mohammed V, Rabat;$^{(f)}$Institute of Applied Physics, Mohammed VI Polytechnic University, Ben Guerir; Morocco.\\
% ...
$^{44}$Physics Department, Southern Methodist University, Dallas TX; United States of America.\\
% ...
$^{48}$Fakult\"{a}t Physik, Technische Universit{\"a}t Dortmund, Dortmund; Germany.\\
% ...
$^{54}$II. Physikalisches Institut, Georg-August-Universit\"{a}t G\"ottingen, G\"ottingen; Germany.\\
% ...
$^{83}$$^{(a)}$Khalifa University of Science and Technology, Abu Dhabi;$^{(b)}$New York University Abu Dhabi, Abu Dhabi;$^{(c)}$United Arab Emirates University, Al Ain;$^{(d)}$University of Sharjah, Sharjah; United Arab Emirates.\\
% ...
$^{102}$CPPM, Aix-Marseille Universit\'e, CNRS/IN2P3, Marseille; France.\\
% ...
$^{121}$Homer L. Dodge Department of Physics and Astronomy, University of Oklahoma, Norman OK; United States of America.\\
% ...
$^{154}$Raymond and Beverly Sackler School of Physics and Astronomy, Tel Aviv University, Tel Aviv; Israel.\\
% ...
$^{a}$ Also at Affiliated with an institute formerly covered by a cooperation agreement with CERN.\\
$^{b}$ Also at An-Najah National University, Nablus; Palestine.\\
$^{c}$ Also at Borough of Manhattan Community College, City University of New York, New York NY; United States of America.\\
$^{d}$ Also at Center for Interdisciplinary Research and Innovation (CIRI-AUTH), Thessaloniki; Greece.\\
$^{e}$ Also at Centre of Physics of the Universities of Minho and Porto (CF-UM-UP); Portugal.\\
$^{f}$ Also at CERN, Geneva; Switzerland.\\
% ...

\end{flushleft}


%

%
%

%

%
%
%
%

\end{document}